\documentclass[english,twocolumn,aps,pra,epsfig,graphics,showpacs]{revtex4}
\usepackage{amsmath}
\usepackage{graphicx}
\usepackage{amssymb}

\makeatletter \providecommand{\tabularnewline}{\\} \makeatletter
\usepackage{amsfonts}
\usepackage{bbm}
\newcommand{\bs}[1]{ \boldsymbol{ #1} }
\makeatother
\usepackage{babel}
\makeatother
\begin{document}

\title{Molecular Dipolar Crystals as High Fidelity Quantum Memory for Hybrid Quantum Computing}

\author{P. Rabl and P. Zoller}

\affiliation{Institute for Theoretical Physics, University of Innsbruck, and\\
 Institute for Quantum Optics and Quantum Information of the Austrian
Academy of Science, 6020 Innsbruck, Austria\\
 }

\begin{abstract}
We study collective excitations of rotational and spin states of an ensemble of polar molecules,
which are prepared in a dipolar crystalline phase, as a candidate for a high fidelity quantum
memory. While dipolar crystals are formed in the high density limit of cold clouds of polar
molecules under 1D and 2D trapping conditions, the crystalline structure protects the molecular
qubits from detrimental effects of short range collisions. We calculate the lifetime of the quantum
memory by identifying the dominant decoherence mechanisms, and estimate their effects on gate
operations, when a molecular ensemble qubit is transferred to a superconducting strip line cavity
(circuit QED). In the case rotational excitations coupled by dipole-dipole interactions we identify
phonons as the main limitation of the life time of qubits.  We study specific setups and
conditions,  where the coupling to the phonon modes is minimized. Detailed results are presented
for a 1D dipolar chain.
\end{abstract}


\maketitle

\section{Introduction}

\label{sec:Intro}
\begin{figure}[t]
\begin{centering}
\includegraphics[width=0.47\textwidth]{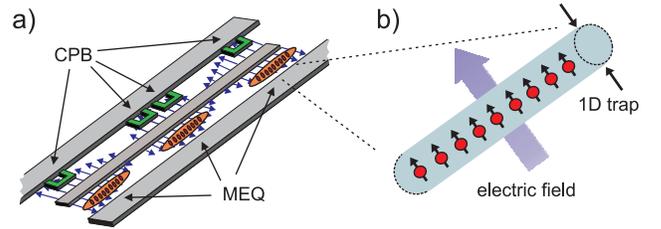}
\caption{a) Schematic picture of a hybrid quantum computer consisting of Cooper Pair Boxes (CPB)
representing a solid state quantum processes and molecular ensemble qubits (MEQ) acting as
long-lived quantum memories. Quantum information is shuttled between to two systems via a
superconducting strip line cavity. b) Dipolar crystal of polar molecules. Under 1D (or 2D) trapping
conditions and dipole moments aligned by a strong electric field, repulsive dipole-dipole
interactions stabilize the molecules against short-range collisions.  See text for more details.}
\label{fig:MDCSetup}
\par\end{centering}
\end{figure}

In recent publications~\cite{singlemol,molensemble} we have studied trapped polar
molecules strongly coupled to a superconducting microwave strip line cavity, which represents a
basic building block for hybrid quantum circuits, interfacing high-fidelity molecular quantum
memory with solid state elements such as Cooper pair boxes
(CPB)~\cite{ChargeQubitRev,Vion2002,Nakamura2003,circuitQED}, superconducting flux
qubits~\cite{Mooij2004} or quantum dots~\cite{LossDot,Marcus2005,Vandersypen2006,DotCavity}. This
suggests a hybrid quantum computing scenarios with the goal of combining the advantages of quantum
optical and solid state implementations by interfacing molecular and solid state qubits in
compatible experimental setups~\cite{TianSorensen2004}. Polar molecules provide two key features
for these interfaces. First, the long coherence times for qubits stored in polar molecules is based
on identifying long-lived rotational or electron and nuclear spin states in the electronic and
vibrational ground state manifold. Second, the strong coupling of molecular qubits to the microwave
cavity is based on transitions between rotational excitations (in the few GHz domain), with large
\emph{electric} dipole moments of the order of a few Debye. In this context, Ref.~\cite{singlemol}
has studied the storage of single polar molecules on electric molecular chips $\sim100$ nm above a
superconducting strip line cavity, resulting in a single molecule - single photon coupling $g$ of
up to 1 MHz. Here, the strong coupling to the microwave cavity provides the additional features of
cooling of the center-of-mass motion of molecules and read out of the molecular qubits. In Ref.
\cite{molensemble} a setup was studied (see Fig.~\ref{fig:MDCSetup}), where a qubit was stored in
the collective spin (or rotational) excitation of a thermal ensemble of $N$ cold polar molecules,
with the advantage of an enhanced collective coupling $g_{N}\equiv g\sqrt{N}$ of the qubit to the
cavity (typically of the order of $1$ to $10$ MHz for $N=10^{4}-10^{6}$ and trapping distances
$\sim10\mu m$), but at the expense of introducing dephasing of the qubit due to state dependent
collisions.

In the present work we will investigate a molecular ensemble quantum memory in the form of a
\emph{dipolar (self-assembled) crystal}. Formation of a molecular dipolar crystal (MDC) is based on
inducing electric dipole moments $\mu_{ind}$ of polar molecules with an external DC field, which
gives a $\mu_{ind}^{2}/4\pi\epsilon_{0}r^{3}\equiv$ $C_{3}/r^{3}$ interaction between molecules at
distance $r$. For molecules confined to 1D or 2D by an external trapping potential and dipole
moments aligned perpendicular, this interaction is repulsive and allows the formation of a
\emph{high-density} crystal, where molecules perform small oscillations around their equilibrium
positions, reminiscent of a Wigner crystal of trapped ions~\cite{WignerRev}. In this crystalline
phase close encounter collisions are strongly suppressed in comparison with thermal ensembles.

The paper is organized as follows. We start in Sec.~\ref{sec:Intro2} with a brief review on
molecular ensemble qubits and the hybrid quantum processor proposed in Ref.~\cite{molensemble}. In
Sec.~\ref{sec:Rotation} we then study the dynamics of collective rotational excitations in a MDC
due to state dependent dipole-dipole interactions and the coupling to motional degrees of freedom.
We derive a general model for this system and calculate the resulting limitations for the lifetime
of ensemble qubits stored in rotational degrees of freedom. In Sec.~\ref{sec:Spin} we extend our
model to molecules with an additional spin degree of freedom and show how we can realize a highly
protected quantum memory using spin ensemble qubits in a MDC. In Sec.~\ref{sec:Specific} we discuss
in some detail a potential experimental implementation of a MDC using electrostatic traps, where in
particular we study the effects of an additional longitudinal confining potential. A summary and
concluding remarks are given in Sec.~\ref{sec:Summary}.


\section{Overview and Background Material}

\label{sec:Intro2}Before introducing our models of ensemble qubits
in dipolar crystals in detail we find it useful to summarize briefly
ensemble qubits and their coupling to microwave cavities, and dipolar
crystals, to introduce and motivate the models for rotational and
spin qubits described in the following sections.

\subsection{Molecular ensembles coupled to a superconducting strip line cavity}

We consider the setup of Fig.~\ref{fig:MDCSetup} a), where a molecular ensemble is coupled to a
superconducting microwave cavity. In addition, the cavity could be coupled strongly to a CPB (or a
quantum dot) representing a circuit QED system \cite{circuitQED}. As discussed in detail in
Sec.~\ref{sec:RotSpectroscopy}, molecular spectroscopy allows us to identify long-lived rotational
states $|g\rangle$ and $|e\rangle$ within the electronic and vibrational ground state manifold. We
assume that the ensemble of $N$ polar molecules is prepared initially in the state $|g_{1}\dots
g_{N}\rangle$ and is coupled to a single mode of a superconducting microwave cavity with a
frequency $\omega_c$ close to the rotational transition frequency $\omega_{eg}$. The dipole
coupling between the molecules and the cavity can then be written in the form
\begin{equation}
\begin{split}H_{{\rm cav-mol}}=\hbar g\sqrt{N}(R_{e}^{\dag}c+c^{\dag}R_{e})\,,\end{split}
\label{eq:cav-mol}\end{equation}
 with $g$ the single molecule vacuum Rabi frequency and $c$ ($c^\dag$) the cavity annihilation (creation) operator.
The collective molecular operator $R_{e}^{\dag}$ creates symmetric Dicke excitations
$|n_{e}\rangle$ with the lowest two states
\begin{eqnarray*}
|0_{e}\rangle & = & |g_{1}\dots g_{N}\rangle,\\
|1_{e}\rangle & \equiv & R_{e}^{\dag}|0_{e}\rangle=1/\sqrt{N}\sum_{i}|g_{1}\dots e_{i}\dots
g_{N}\rangle,\end{eqnarray*}
 representing an ensemble qubit, in addition to higher excitations
of the form $|2_{e}\rangle=1/\sqrt{2}(R_{e}^{\dag})^{2}|0_{e}\rangle$, \emph{etc}. For low number
of rotational excitations the operator $R_{e}$ fulfills approximate bosonic commutation relations
$[R_{e},R_{e}^{\dag}]\simeq1$. As noted above, for typical experimental parameters the
\emph{collectively enhanced} coupling strength $g_N\equiv g\sqrt{N}$ can be of the order of ten
MHz, exceeding experimentally demonstrated decay rates of high-Q superconducting strip line
cavities by several orders of magnitude~\cite{HighQ1,HighQ2}.

For a single molecular ensemble coupled to a single CPB the total Hamiltonian for the hybrid system
is
\begin{equation}\label{eq:Hsys}
\begin{split}
H_{\rm sys}=&\,H_{\rm CPB}+ \hbar \omega_c c^\dag c + \hbar \omega_{eg} R_e^\dag R_e \\
 &+\hbar g_c(\sigma_+ c + \sigma_- c^\dag)  + \hbar g_N(R_{e}^{\dag}c+c^{\dag}R_{e})\,.
\end{split}
\end{equation}
Here terms in the first line represent the bare Hamiltonian operators for the CPB, the cavity and
the molecular ensemble. The second line of Eq.~\eqref{eq:Hsys} contains the Jaynes-Cummings type
interaction between cavity and CPB  with a vacuum Rabi frequency $g_c$ (see Ref.~\cite{circuitQED})
as well as the molecule-cavity interaction given in Eq.~\eqref{eq:cav-mol}. Interactions can in
principle be controlled by tuning frequencies in and out of resonance or, as we explain later in
this paper (see Sec.~\ref{sec:Gates}), by using a switchable Raman process.
Hamiltonian~\eqref{eq:Hsys} is readily generalized to multiple molecular ensembles and CPBs.

In summary Hamiltonian~\eqref{eq:Hsys} provides the basic ingredient to convert the state of the
CPB into a cavity photon superposition state
$|\psi_{c}\rangle=\alpha|0_{c}\rangle+\beta|1_{c}\rangle$~\cite{Yale:StateTransfer}, and in a
second step to map it on an ensemble superposition  \begin{equation}
(\alpha|0_{c}\rangle+\beta|1_{c}\rangle)|0_{e}\rangle\leftrightarrow|0_{c}\rangle(\alpha|0_{e}\rangle+\beta|1_{e}\rangle)\,,
\end{equation}
and vice versa. As discussed in Ref.~\cite{molensemble} coupling of a molecular ensemble qubit to
higher excitations $|2_{e}\rangle$ \emph{etc}. can be suppressed by employing the CPB as a tool  to
generate single photons for state preparation and as a non-linear phase shifter to construct two
qubit gates between different molecular ensembles.

For molecular ensembles to qualify as a quantum memory we not only need fast read/write operations
but we also demand that the lifetime of an arbitrary ensemble superposition
$|\psi_{e}\rangle=\alpha|0_{e}\rangle+\beta|1_{e}\rangle$ is substantially longer than coherence
times of solid state qubits. The lifetime of an ensemble qubits is limited by the single molecule
decoherence time as well as mutual interactions. With expected single molecule decoherence rates of
the order of a few Hz~\cite{singlemol} we identify state dependent elastic and inelastic
collisions~\cite{molensemble} between molecules as the fundamental limitation of the lifetime of a
molecular ensemble quantum memory stored in a thermal gas of molecules.

\subsection{Dipolar crystal }

To avoid collisional dephasing while still keeping the benefit from a collectively enhanced
coupling we consider molecular ensemble qubits prepared in a high density crystalline phase under
1D or 2D trapping conditions. Dipolar crystals of molecules have recently been studied in
Ref.~\cite{2DBuechler}, where it has been shown that with dipole moments aligned by an external DC
electric field molecules are stabilized by repulsive dipole-dipole interactions
$V_{dd}=C_{3}/r^{3}$ in a plane perpendicular to the bias field. Attractive interactions along the
remaining direction are suppressed by a strong transverse confinement
\cite{2DBuechler,MicheliConfinement}. Under such conditions the low temperature physics of the
molecules is characterized by the dimensionless parameter \begin{equation} \gamma=\frac{E_{{\rm
pot}}}{E_{{\rm
kin}}}\equiv\frac{\mu_{ind}^{2}/(4\pi\epsilon_{0}a_{0}^{3})}{\hbar^{2}/ma_{0}^{2}}=\frac{C_{3}m}{\hbar^{2}a_{0}}\,,\label{eq:gamma}\end{equation}
 which is the ratio between potential energy and kinetic energy for
molecules of mass $m$ for a given density $n=1/(a_{0})^{d}$ and dimension $d=1,2$. For $\gamma\gg1$
the dipolar repulsion wins over kinetic energy leading to the formation of a crystalline phase,
i.e. small oscillations of the molecules around their equilibrium values. The formation of a
dipolar crystal at \emph{high} densities, i.e. in the limit where collisions are most damaging is
in contrast to the familiar Wigner crystal for trapped ions or electrons where crystallization
occurs at \emph{low} densities. Numerical Monte Carlo
simulations~\cite{1DLozovik,1DCitro,2DBuechler,2DLozovik} have predicted a crystalline phase for
$\gamma\geq1$ in 1D and $\gamma\geq20$ in 2D. For typical experimental numbers a stable crystal of
polar molecules is found for a lattice spacing $a_{0}$ of a few times 100 nm.

\subsection{Ensemble qubits in dipolar crystals}

  In the following sections we address the question whether it
is possible to achieve a stable molecular crystal \emph{and} at the same time encode quantum
information in ensemble superpositions $|\psi_{e}\rangle=\alpha|0_{e}\rangle+\beta|1_{e}\rangle$.
In this context we distinguish between two types of ensemble qubits: (i) rotational qubits, i.e.
the states $|0_{e}\rangle$ and $|1_{e}\rangle$ introduced above, and (ii) spin qubits. In
Sec.~\ref{sec:Rotation} we first consider rotational ensemble qubits which are directly affected by
state dependent dipole-dipole interactions and decay by phonon induced scattering processes out of
the symmetric state $|1_e\rangle$.
In Sec.~\ref{sec:Spin} we extend our model to molecules with an additional spin degree of freedom
and study ensemble qubits encoded in collective excitations of two spin states $|g\rangle$ and
$|s\rangle$ within the same rotational manifold. As spin degrees of freedom are essentially
unaffected by dipole-dipole interactions, spin ensemble qubits in a MDC form indeed a highly
protected quantum memory. However, a degrading of the spin ensemble quantum memory due to
dipole-dipole interactions still occurs during gate operations when molecules are (virtually)
excited into the rotational state $|e\rangle$. To estimate the resulting gate fidelities under
realistic experimental conditions we focus in Sec.~\ref{sec:Specific} on a specific implementation
of a 1D dipolar crystal and include effects of an additional longitudinal confining potential into
our model.

\section{Rotational Ensemble Qubits in a homogeneous Dipolar Crystal}

\label{sec:Rotation}

In this section we consider the properties of ensemble qubits with $N$ molecules prepared in a
crystalline phase, and qubits encoded in collective rotational excitations. Our goal is to study
the dynamics, and thus decoherence, of an initial rotational ensemble qubit
$|\psi_{e}\rangle=\alpha|0_{e}\rangle+\beta|1_{e}\rangle$ under the influence of dipole-dipole
interactions. We start with the simplest possible model of a homogeneous 1D or 2D crystal,
returning to questions of experimental implementations and requirements (e.g. questions of
transverse and longitudinal trapping potentials, and the choice of particular molecular states) at
a later stage.


\subsection{Hamiltonian}

Let us consider an {\em (infinite) homogeneous} dipolar crystal of a given density corresponding to
a lattice spacing $a_0$, which is initially prepared in the qubit state $|0_{e}\rangle=|g_{1}\dots
g_{N}\rangle$. We denote by ${\bf r}_{i}^{0}$ the classical equilibrium positions of the molecules,
which form a linear chain in 1D or a triangular lattice in 2D. As discussed in
Sec.~\ref{sec:Intro2}, the stability of the crystal requires $\gamma=C_{3}m/\hbar^{2}a_{0}\gg1$,
where now $C_{3}=\mu_{g}^{2}/4\pi\epsilon_{0}$ is determined by $\mu_{g}$, the induced dipole
moment of state $|g\rangle$. The dynamics of the system including internal and motional degrees of
freedom is given by the Hamiltonian \begin{equation} H_{{\rm MDC}}=\sum_{i}\left(\frac{{\bf
p}_{i}^{2}}{2m}+\hbar\omega_{eg}|e_{i}\rangle\langle e_{i}|\right)+\hat{V}_{dd}(\{{\bf
r}_{i}\})\,,\label{eq:MDC0}\end{equation}
 with 1D or 2D position and momentum operators denoted by ${\bf r}_{i}$
and ${\bf p}_{i}$, respectively, and $\omega_{eg}$ the transition frequency between states
$|e\rangle$ and $|g\rangle$. The Hamiltonian \eqref{eq:MDC0} is the sum of the kinetic energies of
the molecules, a bare molecular Hamiltonian for the internal (rotational) states, and the
dipole-dipole interaction $\hat{V}_{dd}(\{{\bf r}_{i}\})$ which couples the internal and motional
degrees of freedom. With $\bs{\mu}$ denoting the electric dipole operator of the molecule the
dipole-dipole interaction is given by
\begin{equation} \hat{V}_{dd}(\{{\bf
r}_{i}\})=\frac{1}{8\pi\epsilon_{0}}\sum_{i\neq j}\frac{\bs{\mu}_{i}\!\cdot\!\bs{\mu}_{j}-3({\bf
n}_{ij}\!\cdot\!\bs{\mu}_{i})({\bf n}_{ij}\!\cdot\!\boldsymbol{\mu}_{j})}{|{\bf r}_{i}-{\bf
r}_{j}|^{3}}\,,\label{eq:Vdd}\end{equation}
 where ${\bf n}_{ij}={\bf r}_{ij}/|{\bf r}_{ij}|$ is the unit vector pointing
along direction ${\bf r}_{ij}={\bf r}_{i}-{\bf r}_{j}$.

To study the dynamics of ensemble states $|0_{e}\rangle$ and $|1_{e}\rangle$ under the action of
$H_{{\rm MDC}}$ we proceed as follows. Since in the crystalline phase molecules are located around
equilibrium positions ${\bf r}_{i}\approx{\bf r}_{i}^{0}$ we describe in a first step the action of
the internal operator $\hat{V}_{I}\equiv\hat{V}_{dd}(\{{\bf r}_{i}^{0}\})$ on the qubit state
$|1_{e}\rangle$. As the dipole-dipole interactions depend on the actual choice of rotational states
$|g\rangle$ and $|e\rangle$ we  start with a short overview on rotational spectroscopy of polar
molecules in Sec.~\ref{sec:RotSpectroscopy}. In Sec.~\ref{sec:DipoleDipoleInteractions} we then
discuss the action of dipole-dipole interactions on ensemble states $|0_{e}\rangle$ and
$|1_{e}\rangle$. We find that – at least in a homogeneous crystal – the only effect of
$\hat{V}_{I}$ is an energy shift for state $|1_{e}\rangle$ which does not destroy the decoherence
of a qubit state $|\psi_{e}\rangle$. Therefore, in Sec.~\ref{sec:FullModel} we include molecular
motion and write the position operator of each molecule as ${\bf r}_{i}={\bf r}_{i}^{0}+{\bf
x}_{i}$, with ${\bf x}_{i}$ accounting for small fluctuations around the classical equilibrium
positions ${\bf r}_{i}^{0}$. By expanding $\hat{V}_{dd}(\{{\bf r}_{i}\})$ in powers of ${\bf
x}_{i}$ we obtain the dominant contributions for the interactions between internal and external
degrees of freedom which, for example, account for state dependent forces on the molecules due to a
difference of the induced dipole moments of states $|g\rangle$ and $|e\rangle$. The analysis of the
resulting model given in Sec.~\ref{sec:1D2DCrystal} finally allows us to estimate the ensemble
quantum memory lifetime, $T_{e}$, for a wide range of system parameters.

\subsection{Rotational Spectroscopy }

\label{sec:RotSpectroscopy}Since the interactions between molecules depend on the actual choice of
rotational states $|g\rangle$ and $|e\rangle$ we will first summarize the rotational spectroscopy
of polar molecules in the presence of external electric fields~\cite{MolBooks}. To keep the
discussion on a basic level we consider in this section only molecules like SrO or CsRb with a
closed electron shell and a $^{1}\Sigma$ electronic ground state. In Sec.~\ref{sec:Spin} we extend
our model to molecules with additional spin degrees of freedom.

At sub-Kelvin temperatures with electronic and vibrational degrees
of freedom frozen out the energy spectrum of a $^{1}\Sigma$ molecule
is well described by the rigid rotor Hamiltonian $H_{M}=B{\bf N}^{2}$
with ${\bf N}$ the angular momentum of the nuclei and $B$ the rotational
constant which is typically in the order of several GHz. In the presence
of an external electric bias field ${\bf E}_{b}$, polar molecules
interact with the field via the dipole coupling $-\boldsymbol{\mu}{\bf E}_{b}$,
with $\boldsymbol{\mu}$ the electric dipole operator of the molecule.
In the following we choose our z-axis along the direction of the bias
field, i.e. ${\bf E}_{b}=E_{b}{\bf e}_{z}$ and the total Hamiltonian
is $H_{M}=B{\bf N}^{2}-\mu_{z}E_{b}$.

\begin{figure}[t]
\begin{centering}
\includegraphics[width=0.47\textwidth]{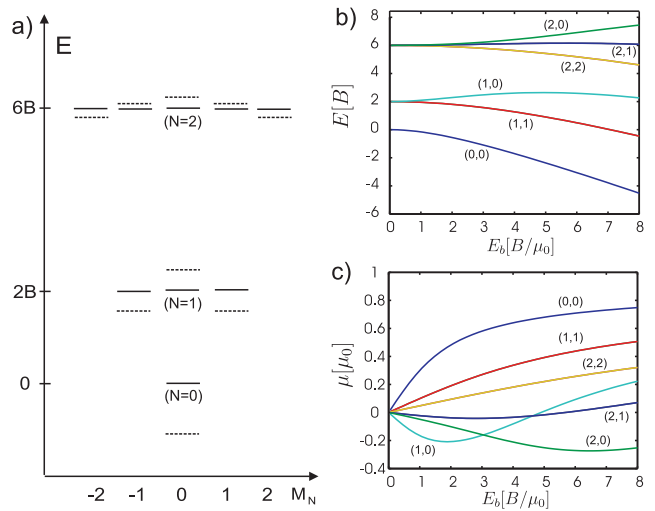}
\caption{a) Rotational energy spectrum for zero bias field $E_b=0$ (solid lines) and
the corresponding shifted energy levels at $E_b=3B/\mu_0$ (dashed lines). b) Energies spectrum and
c) induced dipole moments $\mu_i=\langle i|\mu_z|i\rangle$ for the lowest few eigenstates as a
function of the applied bias field $E_b$. Different curves for eigenstates $|N,M_N\rangle_{E_b}$
are labelled by $(N,|M_N|)$.} \label{fig:RotQubits}
\par\end{centering}
\end{figure}

For a given electric field we label eigenstates of $H_{{\rm M}}$ as $|N,M_{N}\rangle_{E_{b}}$ with
$M_{N}$ the eigenvalue of the operator $N_{z}$. In the field free case ($E_{b}=0$) the eigenstates
$|N,M_{N}\rangle_{0}$ are the usual angular momentum eigenfunctions and the corresponding
anharmonic energy spectrum, $E_{N}=B(N+1)N$, with energy levels $(2N+1)$-fold degenerate is shown
in Fig.~\ref{fig:RotQubits} a). In the presence of an electric bias field ($E_{b}\neq0$) the dipole
coupling mixes different angular momentum eigenfunctions and general eigenstates
$|N,M_{N}\rangle_{E_{b}}$ are superpositions of states $|N,M_{N}\rangle_{0}$ with different $N$ but
with the same $M_{N}$ quantum number. The modified rotor spectrum as a function of the applied
electric field $E_{b}$ is show in Fig.~\ref{fig:RotQubits} b). The spectrum retains its
anharmonicity over a wide range of electric fields values and only for $E_{b}\gg B/\mu_{0}$ with
$\mu_0$ the axis fixed dipole moment of the molecule rotational excitations become approximately
harmonic.

The mixing of different rotational states $|N,M_{N}\rangle_{0}$ in the presence of an electric
field corresponds to an alignment of the molecules along the field direction, and in general an
eigenstate $|\phi\rangle=|N,M_{N}\rangle_{E_{b}}$ exhibits a finite induced dipole moment
$\langle\phi|\mu_{z}|\phi\rangle\neq0$ along the z-direction. The magnitude and sign of the induced
dipole moments depend on the specific state and the strength of the bias field $E_{b}$. The induced
dipole moments for the lowest rotational states are plotted in Fig.~\ref{fig:RotQubits} c) as a
function of the electric field $E_{b}$. We find that for an axis fixed dipole moment
$\mu_{0}\approx5-10$ Debye and moderate electric fields, $E_{b}\approx2B/\mu_{0}$, which typically
corresponds to a few kV/cm, induced dipole moments of a few Debye can be achieved. Note that in
general induced dipole moments of non-degenerate states are different, although there are certain
values of the bias field, so-called `sweet spots'~\cite{singlemol}, where induced dipole moments
for two states are the same.

The anharmonicity of the rotor spectrum for low and moderate electric fields allows us to choose
two rotational states, e.g. $|g\rangle=|N,M_{N}\rangle_{E_{b}}$ and
$|e\rangle=|N',M'_{N}\rangle_{E_{b}}$, which are selectively coupled by a dipole transition to the
fields of a microwave cavity of appropriate frequency and polarization. Selection rules for dipole
transitions require $\Delta M_{N}=0,\pm1$, while the additional restriction $\Delta N=\pm1$ applies
only for vanishing $E_{b}$. In the following the states $|g\rangle$ and $|e\rangle$ form our
truncated single molecule basis which defines our ensemble states $|0_{e}\rangle$ and
$|1_{e}\rangle$, as introduced in Sec.~\ref{sec:Intro}.

\subsection{Dipole-dipole interactions}

\label{sec:DipoleDipoleInteractions}

So far we have discussed ensemble qubits on the level of independent molecules. As the crystalline
phase is stabilized by mutual interactions between molecules, which for different internal states
depend, e.g., on the different induced dipole moments, we proceed to study dipole-dipole
interactions given in Eq.~\eqref{eq:Vdd}.

Consider the action of $\hat{V}_{dd}$ on the qubit states. For fixed positions ${\bf r}_{i}$ we
define the operator acting on the internal states, \begin{equation} \hat{D}_{ij}({\bf
n}_{ij})=\bs{\mu}_{i}\!\cdot\!\bs{\mu}_{j}-3({\bf n}_{ij}\!\cdot\!\bs{\mu}_{i})({\bf
n}_{ij}\!\cdot\!\boldsymbol{\mu}_{j})\,.\label{eq:Dijfull}\end{equation}
 We will simplify $\hat{D}_{ij}({\bf n}_{ij})$ under the assumption
that its action is confined to the two molecule subspace
$\mathcal{H}_{ij}=\{|\epsilon_{i},\epsilon_{j}\rangle,\epsilon_{i}=g,e\}$ and that it is
independent of ${\bf n}_{ij}$ when molecules are confined in the $(x,y)$-plane. This simplification
is possible under the following conditions. First, due to the comparatively large separation
between molecules of $|{\bf r}_{ij}|\gtrsim100$ nm the dipole-dipole interaction is a weak
perturbation on the scale of the rotational spectrum ($\sim B$). This is true for most choices of
states $|g\rangle$ and $|e\rangle$ and allows us to restrict the action of the operator
$\hat{D}_{ij}({\bf n}_{ij})$ to the subspace $\mathcal{H}_{ij}$. There are exceptions, e.g. the
combination $|g\rangle=|N,0\rangle$ and $|e\rangle=|N',+1\rangle$ where the degeneracy between the
states $|N',1\rangle$ and $|N',-1\rangle$ would lead to resonant transitions out of
$\mathcal{H}_{ij}$. We can nevertheless include those combinations of states in our discussion when
we assume that the degeneracy is lifted, e.g., by additional AC microwave fields. Our second
assumption is that with molecular motion restricted to the $(x,y)$-plane, i.e., ${\bf
n}_{ij}\perp{\bf e}_{z}$ the operator $\hat{D}_{ij}({\bf n}_{ij})$ is independent of ${\bf
n}_{ij}$. This condition is fulfilled for $|g\rangle$ and $|e\rangle$ being eigenstates of $N_{z}$.
For other states, e.g. $|e\rangle$ as eigenstate of the operator $N_{x}$, the directional
dependence of $\hat{D}_{ij}({\bf n}_{ij})$ would lead to an additional $x/y$ dependence in the
models for 2D crystals derived below. For simplicity we ignore these cases in the following
discussion.

With these assumption we decompose $\hat{D}_{ij}=\hat{D}_{ij}^{(i)}+\hat{D}_{ij}^{(r)}$. The first
term, $\hat{D}_{ij}^{(i)}$, describes a state dependent interactions due to a difference in the
induced dipole moments, $\delta\mu=\mu_{e}-\mu_{g}$, with $\mu_{i}=\langle i|\mu_{z}|i\rangle$. It
is diagonal in the qubit basis and we can write it as \begin{equation}
\begin{split}\hat{D}_{ij}^{(i)}= & \left(\mu_{g}\mathbbm{1}_{i}+\delta\mu|e_{i}\rangle\langle e_{i}|\right)\left(\mu_{g}\mathbbm{1}_{j}+\delta\mu|e_{j}\rangle\langle e_{j}|\right)\,.\end{split}
\label{eq:Di}\end{equation}
 The second term, $\hat{D}_{ij}^{(r)}$, accounts for resonant exchange
of rotational quanta between molecules. Introducing a Pauli operators
notation $\sigma_{i}^{+}=|e_{i}\rangle\langle g_{i}|$ \emph{etc.},
it is given by \begin{equation}
\hat{D}_{ij}^{(r)}=D_{r}\left(\sigma_{i}^{+}\sigma_{j}^{-}+\sigma_{i}^{-}\sigma_{j}^{+}\right)\,,\label{eq:Dr}\end{equation}
 with the matrix element $D_{r}=\eta|\langle e|\bs{\mu}|g\rangle|^{2}$,
where $\eta=1$ for rotational states of equal quantum number $M_{N}$
and $\eta=-1/2$ for $M_{N}'=M_{N}\pm1$.

While the diagonal operator $\hat{D}_{ij}^{(i)}$ acting on the qubit
states leads to a state dependent energy shift, the operator $\hat{D}_{ij}^{(r)}$
flips the rotational excitations between neighboring molecules. When
motional degrees of freedom are included, both processes result in
state dependent forces on the molecules.

\subsection{Effect of dipole-dipole interactions for fixed lattice positions}

Let us consider the simple situation where the molecules are frozen at lattice positions ${\bf
r}_{i}^{0}$. In this case the dipole-dipole interaction takes on the form \begin{equation}
\hat{V}_{I}=\frac{1}{8\pi\epsilon_{0}}\sum_{i\neq j}\frac{\hat{D}_{ij}}{|{\bf r}_{i}^{0}-{\bf
r}_{j}^{0}|^{3}}\,.\label{eq:VddI}\end{equation}
 In view of $[\hat{V}_{I},R_{e}^{\dag}]\sim R_{e}^{\dag}$ both ensemble qubit states
$|0_{e}\rangle$ and $|1_{e}\rangle$ are eigenstates of $\hat{V}_{I}$. Therefore, apart from a small
energy shift, the internal part of the dipole-dipole interactions does not limit the lifetime of
the ensemble qubit. This statement, of course, ignores inhomogeneity and finite size effects which
depend on the specific experimental setup (see Sec.~\ref{sec:Specific}). However, these
imperfections can in principle be avoided and do not constitute a fundamental restriction to
ensemble quantum memories in dipolar crystals. We conclude that in our model state dependent forces
and the resulting entanglement with motional degrees of freedom is the primary source of
decoherence.

\subsection{Effect of dipole-dipole interactions including motional couplings}

\label{sec:FullModel} We return to the full Hamiltonian $H_{{\rm MDC}}$
given in Eq.~\eqref{eq:MDC0} which includes internal as well as
external degrees of freedom.

\subsubsection{Decomposition of the dipole-dipole interactions acting on internal
and motional degrees of freedom}

With the assumption that the dipole-dipole interaction is confined
to the subspace $\mathcal{H}_{ij}=\{|\epsilon_{i},\epsilon_{j}\rangle,\epsilon_{i}=g,e\}$,
we write \begin{equation}
\hat{V}_{dd}(\{{\bf r}_{i}\})=\frac{\mu_{g}^{2}}{8\pi\epsilon_{0}}\sum_{i\neq j}\frac{\mathbbm{1}_{ij}+\hat{K}_{ij}}{|{\bf r}_{i}-{\bf r}_{j}|^{3}}\,,\label{eq:Vdd3}\end{equation}
 where we introduced the dimensionless operator $\hat{K}_{ij}$ by \begin{equation}
\hat{D}_{ij}\equiv\mu_{g}^{2}\left(\mathbbm{1}_{ij}+\hat{K}_{ij}\right)\,.\label{eq:decomp}\end{equation}
 This decompositions separates $\hat{V}_{dd}$ into a part which is
independent of the internal state and describes purely repulsive interaction
between molecules which stabilize the crystal. All state dependent
properties of $\hat{V}_{dd}$ are contained in the operator $\hat{K}_{ij}$
given by \begin{equation}
\hat{K}_{ij}=\epsilon\left(|e_{i}\rangle\langle e_{i}|+|e_{j}\rangle\langle e_{j}|\right)+\kappa\left(\sigma_{i}^{+}\sigma_{j}^{-}+\sigma_{i}^{-}\sigma_{j}^{+}\right)\,.\label{eq:Kij}\end{equation}
 Here $\epsilon=(\mu_{e}-\mu_{g})/\mu_{g}$ the normalized difference
of induced dipole moments, and $\kappa=D_{r}/\mu_{g}^{2}$ the normalized
coupling constant for resonant exchange processes. Note that in Eq.~\eqref{eq:Kij}
we have omitted the term $\epsilon^{2}|e_{i},e_{j}\rangle\langle e_{i},e_{j}|$
which is negligible for a low number of rotational excitations, as
is the case for our initial ensemble qubit. Therefore, for a given
choice of states $|g\rangle$ and $|e\rangle$ we characterize dipole-dipole
interactions by the induced dipole moment of the ground state $\mu_{g}$,
and the two dimensionless parameters $\epsilon$ and $\kappa$.

We rewrite the molecular position operators as ${\bf r}_{i}={\bf r}_{i}^{0}+{\bf x}_{i}$ and expand
Eq.~\eqref{eq:Vdd3} in ${\bf x}_{i}$, so that the dipole-dipole interaction splits into three
contributions, \begin{equation} \hat{V}_{dd}(\{{\bf r}_{i}\})=\hat{V}_{I}+\hat{V}_{E}(\{{\bf
x}_{i}\})+\hat{V}_{{\rm int}}(\{{\bf x}_{i}\})\,,\label{eq:Vdecomp}\end{equation}
 with $\hat{V}_{I}$ ($\hat{V}_{E}$) acting on internal (external)
degrees of freedom respectively, while $\hat{V}_{{\rm int}}$ contains all remaining terms which
couple the external and the internal dynamics.

\subsubsection{Excitons: rotational excitations hopping on the lattice}

Let us first return to the internal operator $\hat{V}_{I}$ defined in Eq.~\eqref{eq:VddI}, where by
neglecting a global energy shift we can replace operators $\hat{D}_{ij}$ by
$\mu_{g}^{2}\hat{K}_{ij}$. The operator $\hat{K}_{ij}$ given in Eq.~\eqref{eq:Kij} preserves the
total number of molecules in state $|e\rangle$, but allows a propagation of rotational excitations
on the lattice. In the limit of a low number of rotational excitations we diagonalize $\hat{V}_{I}$
by introducing a set of collective operators $R_{{\bf k}}^{\dag}$ defined by \[ R_{{\bf
k}}^{\dag}|0_{e}\rangle\equiv|1_{{\bf k}}\rangle=1/\sqrt{N}\sum_{j}e^{i{\bf kr}_{j}^{0}}|g_{1}\dots
e_{j}\dots g_{N}\rangle,\]
 with a wave vector ${\bf k}$ restricted to the first Brillouin zone
of the lattice. The operators $R_{{\bf k}}$ fulfill (approximate) bosonic commutation relations,
$[R_{{\bf k}},R_{{\bf {k}'}}^{\dag}]\simeq\delta_{{\bf kk}'}$, and in the following we refer to
states created by $R_{{\bf k}}^{\dag}$ as `excitons'. This nomenclature is based on the
similarities of rotational excitations with localized Frenkel excitions in organic crystals
~\cite{OrganicCrystals}. We then identify our qubit state $|1_{e}\rangle$ as associated with the
zero momentum exciton, $R_{e}^{\dag}\equiv R_{{\bf k}=0}^{\dag}$. Including the energy offset
$\hbar\omega_{eg}$ the dynamics of these excitons is given by the Hamiltonian
\begin{equation}\label{eq:Hexc}
H_{{\rm exc}}=\sum_{i}\hbar\omega_{eg}|e_{i}\rangle\langle e_{i}|+\hat{V}_{I}=\sum_{{\bf k}}E({\bf
k})R_{{\bf k}}^{\dag}R_{{\bf k}}\,.\end{equation}
 The energy band of rotational excitations, $E({\bf k})$, is given
in Eq.~\eqref{eq:ExcitonSpectrum} of Sec.~\ref{sec:1D2DCrystal} where we will discuss it in more
detail. For the moment we simply note that Hamiltonian $H_{{\rm exc}}$ is diagonal in ${\bf k}$,
such that $[H_{{\rm exc}},R_{e}^{\dag}]=E(0)R_e^\dag$ as already pointed out at end of
Sec.~\ref{sec:DipoleDipoleInteractions}.

\subsubsection{Phonons }

In a next step we consider the dynamics of external degrees of freedom
of the molecules which is determined by the interaction $\hat{V}_{E}(\{{\bf x}_{i}\})$
and the kinetic energy $H_{{\rm kin}}=\sum_{i}{\bf p}_{i}^{2}/2m$.
Since the first order expansion of $\hat{V}_{E}(\{{\bf x}_{i}\})$
vanishes due to the definition of equilibrium positions ${\bf r}_{i}^{0}$,
the first non-vanishing contribution is of second order in ${\bf x}_{i}$
and given by \begin{equation}
\hat{V}_{E}(\{{\bf x}_{i}\})=\frac{3\mu_{g}^{2}}{16\pi\epsilon_{0}}\sum_{i\neq j}\frac{5\left[({\bf x}_{i}-{\bf x}_{j})\cdot{\bf n}_{ij}^{0}\right]^{2}-\left[{\bf x}_{i}-{\bf x}_{j}\right]^{2}}{|{\bf r}_{i}^{0}-{\bf r}_{j}^{0}|^{5}}\,.\end{equation}
 The quadratic interaction between molecules causes collective oscillations
(phonons) in the crystal described by the Hamiltonian $H_{{\rm phon}}=H_{{\rm kin}}+\hat{V}_{E}(\{{\bf x}_{i}\})$.
As $H_{{\rm phon}}$ is simply a set of coupled harmonic oscillators
it can be written in diagonal form \begin{equation}
H_{{\rm phon}}=\sum_{{\bf q},\lambda}\hbar\omega_{\lambda}({\bf q})a_{\lambda}^{\dag}({\bf q})a_{\lambda}({\bf q})\,.\label{eq:Hphon}\end{equation}
 Here we introduced the annihilation (creation) operators $a_{\lambda}({\bf q})$
($a_{\lambda}^{\dag}({\bf q})$) for phonons of quasi momentum ${\bf q}$
and frequency $\omega_{\lambda}({\bf q})$. In 2D, the index $\lambda$
labels the two different phonon branches. The phonon modes in the
dipolar crystal are acoustic phonons. A discussion of the frequency
spectrum is given in Sec.~\ref{sec:1D2DCrystal}.


\subsubsection{Exciton-Phonon Interactions}

The remaining terms of $\hat{V}_{dd}(\{{\bf r}_{i}\})$ which can
not be decomposed into purely internal or external operators are summarized
in $\hat{V}_{{\rm int}}(\{{\bf x}_{i}\})$. The first non-vanishing
order of $\hat{V}_{{\rm int}}(\{{\bf x}_{i}\})$ is linear in the
operators ${\bf x}_{i}$ and is given by \begin{equation}
\hat{V}_{{\rm int}}(\{{\bf x}_{i}\})\simeq-\frac{3\mu_{g}^{2}}{8\pi\epsilon_{0}}\sum_{i\neq j}
\frac{{\bf r}_{i}^{0}-{\bf r}_{j}^{0}}{|{\bf r}_{i}^{0}-{\bf r}_{j}^{0}|^{5}}\,({\bf x}_{i}-{\bf x}_{j})\otimes\hat{K}_{ij}\,.\label{eq:Vint}
\end{equation}
 It describes a state dependent force on the molecules and entangles
internal and external degrees of freedom. In the following we introduce a new symbol $H_{{\rm
int}}\equiv\hat{V}_{{\rm int}}(\{{\bf x}_{i}\})$ and rewrite Eq.~\eqref{eq:Vint} in terms of
exciton operators $R_{{\bf k}}$ and the phonon operators $a_{{\bf q}}$. We obtain an interaction
Hamiltonian of the form \begin{equation}\label{eq:Hint}
 H_{{\rm int}}=\sum_{{\bf k},{\bf
q},\lambda}M_{\lambda}({\bf q},{\bf k})[a_{\lambda}({\bf q})+a_{\lambda}^{\dag}(-{\bf q})]\,
R_{{\bf k+q}}^{\dag}R_{{\bf k}}\,,\end{equation}
 which describes scattering processes from state $|{\bf k}\rangle$
into state $|{\bf k}+{\bf q}\rangle$ under the absorbtion (emission)
of a phonon of quasi momentum ${\bf q}$ ($-{\bf q}$). We postpone
a discussion of the explicit form of the coupling matrix elements
$M_{\lambda}({\bf q},{\bf k})$ to Sec.~\ref{sec:1D2DCrystal}.

\subsubsection{Summary }

In summary, we have shown that the dynamics of a molecular dipolar crystal given by $H_{{\rm MDC}}$
in Eq.~\eqref{eq:MDC0} contains the three contributions, \[ H_{{\rm MDC}}=H_{{\rm exc}}+H_{{\rm
phon}}+H_{{\rm int}}\,,\]
 with \begin{eqnarray}
H_{\rm exc} & = & \sum_{{\bf k}}E({\bf k})R_{{\bf k}}^{\dag}R_{{\bf k}},\nonumber \\
H_{\rm phon} & = & \sum_{{\bf q},\lambda}\hbar\omega_{\lambda}({\bf q})a_{\lambda}^{\dag}({\bf q})a_{\lambda}({\bf q}),\label{eq:Hcry}\\
H_{{\rm int}} & = & \sum_{{\bf k},{\bf q},\lambda}M_{\lambda}({\bf q},{\bf k})[a_{\lambda}({\bf
q})+a_{\lambda}^{\dag}(-{\bf q})]\, R_{{\bf k+q}}^{\dag}R_{{\bf k}},\nonumber \end{eqnarray}
 which is the (minimal) model which describes the evolution of ensemble
qubits in a self-assembled molecular dipolar crystal. While explicit expressions for the energy
dispersion $E({\bf k})$, the phonon spectrum $\omega_{\lambda}({\bf q})$ and the scattering matrix
elements $M_{\lambda}({\bf q},{\bf k})$ are give in Sec.~\ref{sec:1D2DCrystal} for the 1D and 2D
crystal we first note the general structure of $H_{{\rm MDC}}$. The ensemble operator
$R_{e}^{\dag}$ is an eigenoperator of $H_{\rm exc}$ as well as $H_{\rm phon}$, and therefore, apart
form an energy shift, the first two lines of Eq.~\eqref{eq:Hcry} preserve the coherence of a qubit
superposition $|\psi_{e}\rangle=\alpha|0_{e}\rangle+\beta|1_{e}\rangle$. The third line of
Eq.~\eqref{eq:Hcry}, $H_{\rm int}$, leads to phonon assisted transitions from the symmetric qubit
state $|1_{e}\rangle\equiv|{\bf k}=0\rangle$ into orthogonal states $|{\bf k}\neq0\rangle$. This
loss process is the dominant source of decoherence for a qubit state $|\psi_{e}\rangle$ and in
Sec.~\ref{sec:Lifetime} we calculate the resulting lifetime $T_{e}$ for the ensemble quantum
memory.

\subsection{Molecular dipolar crystals in 1D \& 2D}

\label{sec:1D2DCrystal} In this section we discuss the exciton dispersion $E({\bf k})$, the phonon
spectrum $\omega_{\lambda}({\bf q})$ and the coupling matrix elements $M_{\lambda}({\bf q},{\bf
k})$ which determine the properties of Hamiltonian $H_{{\rm MDC}}$ given in Eq.~\eqref{eq:Hcry}.
For a lattice spacing $a_{0}$ we express those quantities in terms of the dipole-dipole energy
$U_{dd}=\mu_{g}^{2}/(4\pi\epsilon_{0}a_{0}^{3})$ and the dimensionless parameters $\gamma$,
$\epsilon$ and $\kappa$. Our focus is placed on the 1D crystal where we derive analytic expressions
for the relevant quantities. For the 2D crystal we present numerical results and identify the main
differences compared to the 1D case. The derivations of the following results can be found in
App.~\ref{app:ring} and the main results are summarized in Fig.~\ref{fig:1DCrystal} (1D) and
Fig.~\ref{fig:2DCrystal} (2D).

\subsubsection{Excitons }

The energy spectrum of excitons in the dipolar crystal, $E({\bf k})$, contains three contributions,
\begin{equation} E({\bf k})=\hbar\omega_{eg}+U_{dd}[\epsilon J(0)+\kappa J({\bf
k})]\,,\label{eq:ExcitonSpectrum}\end{equation}
 where the dimensionless band structure $J({\bf k})$ is defined in
App.~\ref{app:ring} in Eq.~\eqref{eq:Jfull}. While a finite difference in the dipole moments,
$\epsilon\neq0$, only causes a shift of the transition frequency with $J(0)=2\zeta(3)$ in the 1D
case and $J(0)\simeq11.034$ in 2D, the resonant exchange processes proportional to $\kappa$ lead to
the formation of an exciton band structure as shown in Figs.~\ref{fig:1DCrystal}
and~\ref{fig:2DCrystal}. For the 1D crystal the explicit expression for $J(k)$ is given in
App.~\ref{app:ring} in Eq.~\eqref{eq:J1D} and we find that in the long wavelength limit
$k\rightarrow0$ it exhibits a non-analytic behavior, \begin{equation} J(k)\simeq
J(0)-3/2[1-3/2\log(ka_{0})](ka_{0})^{2},\end{equation}
 which is a consequence of the slow decay of dipole-dipole interactions.
For the 2D crystal the long-range character of dipole-dipole interactions is even more apparent and
results in a linear dispersion $E({\bf k})\!-\! E(0)\sim|{\bf k}|$ for small $|{\bf k}|$. The total
width of the energy band is $\Delta E=7\zeta(3)/2\times|\kappa|U_{dd}$ in 1D and $\Delta
E\simeq13.37\times|\kappa|U_{dd}$ in 2D. Note that for rotational states with $\kappa$ positive the
band structure is `inverted' and the long-wavelength excitations have the highest energy.

\subsubsection{Phonon spectrum }

\begin{figure}
\begin{centering}
\includegraphics[width=0.4\textwidth]{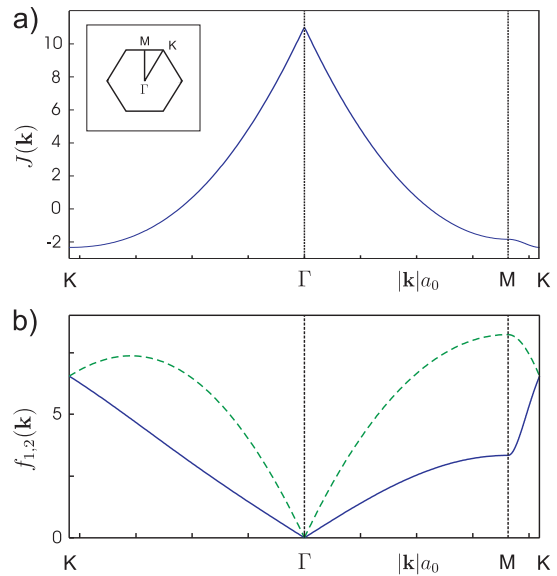}
\caption{Dipolar crystal in 2D: a) Dimensionless band structure $J({\bf k})$ (see
Eq.~\eqref{eq:ExcitonSpectrum}).
 b) Dimensionless phonon spectrum $f_\lambda({\bf q})$ (see
Eq.~\eqref{eq:PhononSpectrum}). Band structure and phonon spectrum are plotted for {\bf k} vectors
along the symmetry lines of the first Brilluoin zone with symmetry points indicated in the inset of
plot a). } \label{fig:2DCrystal}
\par\end{centering}
\end{figure}

As shown in App.~\ref{app:ring} the spectrum of the acoustic phonon modes in the self-assembled
dipolar crystal is of the general from \begin{equation} \hbar\omega_{\lambda}({\bf
q})=\sqrt{\frac{1}{\gamma}}\, U_{dd}\, f_{\lambda}({\bf
q})\,.\label{eq:PhononSpectrum}\end{equation} In the 1D crystal the dimensionless function $f(q)$
defined in Eq.~\eqref{eq:f} has a long wavelength limit $f(q)\simeq\sqrt{12\zeta(3)}\times qa_{0}$,
and a maximum value of $f(\pi)=\sqrt{93\zeta(5)/2}\simeq6.94$. The phonon spectrum is therefore
characterized by the sound velocity $c=\sqrt{12\zeta(3)/\gamma}\, a_{0}U_{dd}/\hbar$ and the Debye
frequency, $\omega_{D}\equiv\omega(\pi)$, \begin{equation}\label{eq:wD}
\hbar\omega_{D}=\sqrt{\frac{93\zeta(5)}{2\gamma}}\times U_{dd}\,.\end{equation}
 The full phonon spectrum is plotted in Fig.~\ref{fig:1DCrystal}
b). In the 2D crystal there are two acoustic phonon branches, $\lambda=1,2$ and the corresponding
dimensionless spectra $f_{\lambda}({\bf q})$ are plotted in Fig.~\ref{fig:2DCrystal} b). In 2D the
maximum phonon frequency is $\hbar\omega_{D}\simeq8.22\times U_{dd}/\sqrt{\gamma}$.

\subsubsection{Exciton-phonon interactions }

\begin{figure}
\begin{centering}
\includegraphics[width=0.45\textwidth]{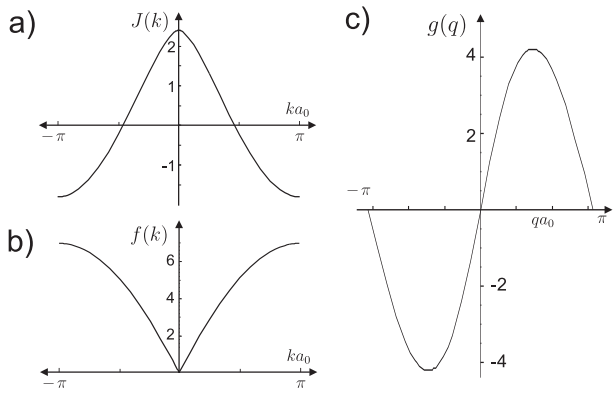}
\caption{Dipolar crystal in 1D: a) Dimensionless band structure $J(k)$ (see
Eq.~\eqref{eq:ExcitonSpectrum}). b) Dimensionless phonon spectrum $f(q)$ (see
Eq.~\eqref{eq:PhononSpectrum}). c) Function $g(q)$ which enters in the expression of coupling
matrix elements  $M(q,k)$ given in Eq.~\eqref{eq:CouplingMatrixElements}.} \label{fig:1DCrystal}
\par\end{centering}
\end{figure}

Excitons and phonons interact via $H_{{\rm int}}$ given in the second line of Eq.~\eqref{eq:Hcry}.
We write the coupling matrix element as \begin{equation}\label{eq:CouplingMatrixElements}
M_{\lambda}({\bf q},{\bf k})\!=i\frac{U_{dd}}{\gamma^{\frac{1}{4}}}\sqrt{\frac{1}{Nf_{\lambda}({\bf
q})}} \left(\epsilon g_{\lambda}({\bf q})\!+\!\kappa[g_{\lambda}({\bf k}\!+\!{\bf q})\!-\!
g_{\lambda}({\bf k})]\right),\end{equation}
 where we introduced an additional dimensionless function $g_{\lambda}({\bf q})$
defined in App.~\ref{app:ring} in Eq.~\eqref{eq:gfull}. For the 1D crystal the explicit expression
of $g(q)$ is given in Eq.~\eqref{eq:g1D} and plotted in Fig.~\ref{fig:1DCrystal} c). The matrix
element $M_{\lambda}({\bf q},{\bf k})$ contains two contributions. The first is proportional
$\epsilon$ and describes a phonon induced (on-site) energy shift of a molecule in state $|e\rangle$
due to a difference in the induced dipole moments. This type of interaction does only depend on the
transferred momentum ${\bf q}$ and is familiar from polaron models discussed in solid state physics
~\cite{PolarModels}. The second contribution proportional to $\kappa$ describes phonon induced
hopping of excitons. The coupling matrix elements for this process also depend on the initial
exciton state $|{\bf k}\rangle$. In the long wavelength limit, $|{\bf q}|\rightarrow0$, both
contributions scales as $\sim\sqrt{|{\bf q}|}$, and scattering events with low momentum transfer
are suppressed.

Without going into the details of $M_{\lambda}({\bf q},{\bf k})$ we point out two properties which
are relevant for the discussion below. First, for long wavelength excitons the total strength of
the exciton phonon interaction is in the order of
\begin{equation} |{\bf k}|\rightarrow0:\qquad \mathcal{O}(H_{{\rm int}})=|\epsilon+\kappa|
U_{dd}\left(\frac{1}{\gamma}\right)^{\frac{1}{4}}\,,\label{eq:OHint}
\end{equation}
 and therefore only weakly suppressed by the parameter $\gamma$.
This means that in general the exciton-phonon interaction has a considerable effect on the dynamics
of a molecular dipolar crystal even deep in the crystalline phase with $\gamma\gg1$. However, for a
specific choice of rotational states and values of the electric bias field  where
$\epsilon+\kappa=0$ is fulfilled, long wave length excitons completely decouple from the phonon
modes. We come back to this point at the end of Sec.~\ref{sec:Lifetime}.

\subsection{Lifetime of rotational ensemble qubits}

\label{sec:Lifetime} Based on the structure of $H_{{\rm MDC}}$ given
in Eq.~\eqref{eq:Hcry} and the discussion of its properties in subsection~\ref{sec:1D2DCrystal},
we finally return to the original question of the lifetime of an ensemble
qubit prepared in a state $|\psi_{e}\rangle=\alpha|0_{e}\rangle+\beta|1_{e}\rangle$.
As already mentioned above we find that the dominant decoherence mechanism
arises form the decay of the state $|1_{e}\rangle$ due to phonon
assisted scattering into orthogonal states $|{\bf k}\neq0\rangle$.
This means we can identify the lifetime of the ensemble quantum memory,
$T_{e}$, with the lifetime of the state $|1_{e}\rangle$.

\subsubsection{Ensemble qubit decay }

In the following we consider the situation where at some initial time $t=0$ the system is prepared
in the state $\rho_{0}=|0_{e}\rangle\langle0_{e}|\otimes\rho_{T}$with $\rho_{T}$ the equilibrium
density operator of the phonon modes for a temperature $T$. At time $t=0^{+}$ we instantaneously
excite the molecules into the qubit state $|1_{e}\rangle$ and calculate probability $P_{e}(t)$ to
find the system in state $|1_{e}\rangle$ after a waiting time $t$. As we are only interested in
times $t$ where $P_{e}(t)$ is still close to 1, we can use second order perturbation theory and
obtain \begin{equation}
\begin{split} & P_{e}(t)\simeq1-\frac{2}{\hbar^{2}}\int_{0}^{t}dt'\int_{0}^{t'}d\tau\,\sum_{{\bf q},\lambda}|M_{\lambda}({\bf q},0)|^{2}\times\\
 & \left[(N(\omega_{\lambda}({\bf q}))+1)\cos(\Omega^{-}({\bf q})\tau)+N(\omega_{\lambda}({\bf q}))\cos(\Omega^{+}({\bf q})\tau)\right].\end{split}
\label{eq:Pe}\end{equation}
 Here $N(\omega)=1/[\exp(\hbar\omega/k_{B}T)-1]$ is the thermal occupation
number for phonons of frequency $\omega$ and $\Omega^{\pm}({\bf q})=[E(0)-E({\bf
q})]/\hbar\pm\omega({\bf q})$. For very short times Eq.~\eqref{eq:Pe} leads to a quadratic decay of
the excited state probability, \begin{equation}
P_{e}(t)\simeq1-W^{2}t^{2}\,,\label{eq:quad}\end{equation}
 with a characteristic rate $W$ defined by
 \begin{equation}
W^{2}=\frac{1}{\hbar^{2}}\sum_{{\bf q},\lambda}|M_{\lambda}({\bf
q},0)|^{2}\big(2N(\omega_{\lambda}({\bf q}))+1\big)\,.\label{eq:quadRate}
\end{equation} For long
times the decay of $P_{e}(t)$ turns into a linear function of $t$, \begin{equation}
P_{e}(t)\simeq1-\Gamma t\,,\end{equation}
 with the decay rate $\Gamma$ given by Fermi's Golden Rule, \begin{equation}
\begin{split}\Gamma=\frac{2\pi}{\hbar^{2}}\sum_{{\bf q},\lambda}|M_{\lambda}({\bf q},0)|^{2} & \Big[\big(N(\omega_{\lambda}({\bf q}))+1\big)\delta(\Omega({\bf q})-\omega_{\lambda}({\bf q}))\\
 & +N(\omega_{\lambda}({\bf q}))\delta(\Omega({\bf q})\!+\!\omega_{\lambda}({\bf q}))\Big].\end{split}
\label{eq:Goldenrule}\end{equation}
 The crossover time $t_{c}$ between the quadratic and the linear regime is
roughly given by $(t_{c})^{-1}\approx{\rm \max}\{\Delta E/\hbar,\omega_{D}\}$, with the exciton
band width $\Delta E$ and the phonon Debye frequency $\omega_{D}$  discussed in
Sec.~\ref{sec:1D2DCrystal}. As long as the associated decay probability $P_c=W^2t_c^2$ is much
smaller than one, the application of Fermi's Golden Rule is valid and we obtain a lifetime
$T_e=1/\Gamma$. We refer to this case as the weak coupling regime.  Otherwise, for $P_c\approx 1$
or in the strong coupling regime, the qubit decay is determined by the quadratic formula given in
Eq.~\eqref{eq:quad} and we identify $T_e=1/W$. At low temperatures we find that when the exciton
band width $\Delta E$ is larger than $\hbar\omega_{D}$,
$P_{c}\approx(\epsilon+\kappa)^{2}/(\kappa^{2}\sqrt{\gamma})$ while in the opposite case
$P_{c}\approx(\epsilon+\kappa)^{2}$.

\subsubsection{Quadratic decay: strong coupling regime}

In the strong coupling regime the initial non-energy conserving transitions out of the qubit state
$|1_e\rangle$ already lead to a strong quadratic reduction of $P_e(t)$. The corresponding rate $W$
defined in Eq.~\eqref{eq:quadRate} can be written as
\begin{equation}
W^{2}|_{d}=\frac{U_{dd}^{2}}{\hbar^{2}}\frac{(\epsilon+\kappa)^{2}}{\sqrt{\gamma}}\times
\mathcal{I}_{d}\left(\tau=\sqrt{\gamma}\,\frac{k_{B}T}{U_{dd}}\right)\,,\end{equation} with $d=1,2$
the dimension of the crystal and $\mathcal{I}_{d}(\tau)$ a numerical integral
\begin{equation} \mathcal{I}_{d}(\tau)=\sum_{\lambda=1}^{d}\int_{BZ}\frac{d^{d}q}{V_{BZ}}\,\frac{|g_{\lambda}({\bf
q})|^{2}}{f_{\lambda}({\bf q})}\left(\frac{2}{e^{f_{\lambda}({\bf
q})/\tau}-1}+1\right)\,.\end{equation} In the two limiting cases this integral behaves as
$\mathcal{I}_{d}(\tau\rightarrow\infty)=\mathcal{O}(1)$ and $\mathcal{I}_{d}(\tau
\rightarrow0)=\mathcal{O}(\tau)$. Neglecting numerical constants arising from the exact evaluation
of $\mathcal{I}_{d}(\tau)$ we can summarize the estimated qubit lifetime $T_{e}$ as
\begin{equation}
(T_{e})^{-1}\approx\frac{U_{dd}}{\hbar}\frac{|\epsilon+\kappa|}{\gamma^{\frac{1}{4}}}\times{\rm
max}\{1,\gamma^{\frac{1}{4}}\sqrt{k_{b}T/U_{dd}}\,\}\,.\end{equation}

\subsubsection{ Fermi's Golden Rule: weak coupling regime  }

In the weak coupling regime only energy conserved transitions with a certain transferred momentum
${\bf q}_{0}$ lead to the decay of $P_e(t)$. According to Eq.~\eqref{eq:Goldenrule} ${\bf q_0}$ is
defined by $\Omega({\bf q}_{0})\pm\omega_{\lambda}({\bf q}_{0})=0$, where the positive (negative)
sign applies for a $\kappa<0$ ($\kappa>0$) corresponding to phonon absorption (emission) processes.
Based on the discussion on the exciton and phonon spectra given in Sec.~\ref{sec:1D2DCrystal} we
distinguish between two cases.

In the first case the resonant k-vector is different from zero, ${\bf q}_{0}\neq0$. Such a
situation occurs 1D for $\kappa/\sqrt{\gamma}\gg1$, when the exciton bandwidth exceeds the phonon
band width, $\Delta E\geq\hbar\omega_{D}$. In 2D, due to the linear long-wavelength limit of
$E({\bf k})$ this situation occurs only for a negligible parameter regime where $\Delta
E\simeq\hbar\omega_{D}$. The decay rate for 1D is
\begin{equation}
\Gamma_{1D}=\frac{U_{dd}}{\hbar}\frac{(\epsilon+\kappa)^{2}}{\kappa\sqrt{\gamma}}\times C(q_{0})\times\left[N(\omega(q_{0}))+\Theta(\kappa)\right]\,,\label{eq:GoldenRule2}
\end{equation}
 with $C(q_{0})=2g^{2}(q_{0})/(f(q_0)|J'(q_{0})+f'(q_0)/\sqrt{\gamma}\kappa|)\lesssim 8$.
 The Heaviside function $\Theta(\kappa)$ takes into account that only for $\kappa>0$ there is a
finite decay rate at $T=0$.

In the second case the resonance condition is only fulfilled for ${\bf q}_{0}=0$. This situation is
in general true for a 2D crystal and in 1D for $\kappa/\sqrt{\gamma}<1$. The decay rate then
depends on the $|{\bf q}|\rightarrow0$ limit of the summand in Eq.~\eqref{eq:Goldenrule}. As
$|M({\bf q},0)|^{2}\sim|{\bf q}|$ vanishes in the long wavelength limit, we only obtain a finite
value for $\Gamma$ in  1D and for a non-zero temperature. The resulting rate is
\begin{equation}
\Gamma_{1D}=(\epsilon+\kappa)^{2}\sqrt{\frac{3\zeta(3)}{4}}\times\sqrt{\gamma}\times\frac{k_{B}T}{\hbar}\,.\label{eq:GoldenRule3}
\end{equation}
 In 2D, due to the additional factor of $|{\bf q}|$ in the density
of states, we obtain $\Gamma_{2D}=0$.

\subsubsection{Discussion: rotational ensemble qubits }

In summary we have found that the lifetime $T_{e}$ of a rotational ensemble quantum memory in a
molecular dipolar crystal is primarily determined by the dipole-dipole energy
$U_{dd}=\mu_{g}^{2}/(4\pi\epsilon_{0}a_{0}^{3})$ and depends further on the relation between the
dimensionless parameters $\gamma$, $\epsilon$ and $\kappa$.

For low temperatures the dipole-dipole energy $U_{dd}$ is limited from below by the condition
$U_{dd}\geq\gamma_{c}^{3}\hbar^{6}/C_{3}^{2}m^{3}$, with $\gamma_{c}$ the minimum value of $\gamma$
which guarantees the stability of the crystal. For an upper bound we use the critical value
$\gamma_{c}=20$ predicted in Ref.~\cite{2DLozovik,2DBuechler} for the 2D crystal. For the nominal
parameters $\mu_{g}=1$ Debye and $m=100$ amu we obtain $U_{dd}/h\geq260\,{\rm kHz}$. As this bound
scales like $\sim1/\mu_{g}^{4}$ choosing molecular states with larger induced dipole moments
$\mu_{g}\sim5$ D the scale of $\Gamma$ can in principle be reduced below the kHz regime. However,
for a system at finite temperature $U_{dd}$ is also bound from below by $k_{B}T$. For a 2D crystal
the melting temperature of a dipolar crystal is $T_M\approx0.09U_{dd}/k_{B}$~\cite{2DMelting},
while in 1D we expect a stable crystal for $k_{B}T\lesssim U_{dd}$ (see Sec.~\ref{sec:Specific}).

For a given $U_{dd}$ the lifetime of the ensemble qubit depends further on the dimensionless
parameters $\kappa$ and $\epsilon$ which in turn depend on the rotational basis states $|g\rangle$
and $|e\rangle$ and the value of the applied bias field $E_b$. In Table~\ref{tab:qubits} we have
listed a few specific choices of rotational states $|g\rangle$ and $|e\rangle$ and the
corresponding values of $\kappa$ and $\epsilon$. For certain states we find a `sweet spot' of the
electric bias field where the induced dipole moments of two rotational states are the same and
$\epsilon$ vanishes. The value of $\kappa$ can be significantly reduced by choosing states
$|g\rangle$ and $|e\rangle$ with a (to a high degree) dipole forbidden transition, e.g. $N'=N\pm2$,
with the cost of also lowering the coupling to the microwave cavity. An ideal situation occurs when
both $\epsilon$ and $\kappa$ are large but exactly cancel each other, e.g. $\epsilon+\kappa=0$. For
such a `magic' configuration the dipole-dipole interactions of the triplet states $|gg\rangle$ and
$|ge\rangle+|eg\rangle$ are exactly the same, which makes a symmetric excitation $|1_e\rangle$
insensitive to phonon-induced fluctuations of dipole-dipole interactions.
 Examples for such phonon-decoupled states are given in row e) and f) of
Table~\ref{tab:qubits}. An additional interesting observation is the absence of energy conserving
transitions in a 2D crystal, which in the weak coupling regime implies $\Gamma_{2D}=0$. A detailed
investigation of this case and corrections due to higher order terms in the interactions
Hamiltonian $H_{\rm int}$ are the subject of future research.

\begin{table}
\begin{centering}
\begin{tabular}{|c|c|c|c|c|c|}
\hline & $|N,M_{N}\rangle$  & $E_{b}[B/\mu_{0}]$  & \,\,\,$\mu_{g}[\mu_{0}]$\,\,\,  &
\,\,\,$\kappa$\,\,\,  & \,\,\,$\epsilon$\,\,\, \tabularnewline \hline \,a)
&$\,\,\,|g\rangle=|1,0\rangle,|e\rangle=|2,0\rangle\,\,\,$  & 3.05  & -0.16  & 10.5  & 0
\tabularnewline \hline \,b)&$\,\,\,|g\rangle=|1,0\rangle,|e\rangle=|3,0\rangle\,\,\,$  & 3.91  &
0.09 & 1  & 0 \tabularnewline \hline \,c)&$|g\rangle=|0,0\rangle,|e\rangle=|1,0\rangle$  & 8  &
0.75 & 0.1 & -0.7 \tabularnewline \hline \,d)&$|g\rangle=|0,0\rangle,|e\rangle=|1,1\rangle$  & 5 &
0.68 & -0.24 & -0.29 \tabularnewline  \hline \,e)&$|g\rangle=|0,0\rangle,|e\rangle=|1,0\rangle$ &
1.44 & 0.39 & 1.51  & -1.51 \tabularnewline \hline
\,f)&$|g\rangle=|1,0\rangle,|e\rangle=|3,0\rangle$ & 3.44 & -0.13 & 0.39  & -0.39 \tabularnewline
\hline
\end{tabular}\caption{Different combination of rotational states $|g\rangle$ and $|e\rangle$ and the
corresponding values of $\epsilon=(\mu_e-\mu_g)/\mu_g$ and $\kappa=D_r/\mu_g^2$. States are
specified by the quantum numbers $|N,M_N\rangle$ and the value of the external bias field, $E_b$.
}\label{tab:qubits}
\par\end{centering}
\end{table}

In conclusion we find that the phonon induced decay of rotational ensemble qubits can cover a wide
range of values and is in the end to a large extend determined by the specific experimental
implementation of this system. When trapping of molecules can be achieved independent of the
rotational state, e.g. in optical or magnetic traps, we can find certain magic configurations in
which rotational ensemble qubits  are completely decoupled from the phonon bath. In other cases,
e.g. for electrostatic traps, we are limited to a certain set of trappable rotational states and
typically we are not able to fulfill the decoupling condition $\epsilon+\kappa=0$. In such a
situation it is not a good choice to encode quantum information in the rotational degree of
freedom.

\section{Spin Ensemble Qubits:}
\label{sec:Spin}

In this section we extend our discussion of molecular ensemble qubits to molecules with spin or
hyperfine states. This additional internal degree of freedom which is not directly affected by
electric dipole-dipole interactions allows us to overcome some of the limitations of purely
rotational qubits. In particular, it is then possible to use the rotational degree of freedom  to
trap molecules in electrostatic potentials and encode quantum information in different spin states,
or in contrast  to use spin states for magnetic trapping and encode quantum information in
rotational degrees of freedom.

\subsection{Rotational spectroscopy of $^2\Sigma$ molecules}\label{sec:SpectroscopySpin}
In general the rotational spectroscopy of polar molecules involves spin-rotation interactions
between the rotation of the nuclei ${\bf N}$ and the spin of unpaired electrons ${\bf S}$  as well
as hyperfine interactions between ${\bf S}$ and the nuclear spin ${\bf I}$~\cite{MolBooks}.
However, to keep the following discussion on a basic level we consider in this paper only the case
of a $^2\Sigma$ molecules with a single unpaired electron with spin $S=1/2$ and no nuclear spin
$I=0$ for which hyperfine interactions are absent. As, to our knowledge, non of the polar molecules
studied in current experiments falls into this category~\cite{doyle2004} we explain below how the
arguments presented for the $^2\Sigma$ molecule can be applied for more complicated molecules with
$I>0$ as long as the hyperfine interactions are small compared to the rotational constant $B$.

\begin{figure}
\begin{centering}
\includegraphics[width=0.4\textwidth]{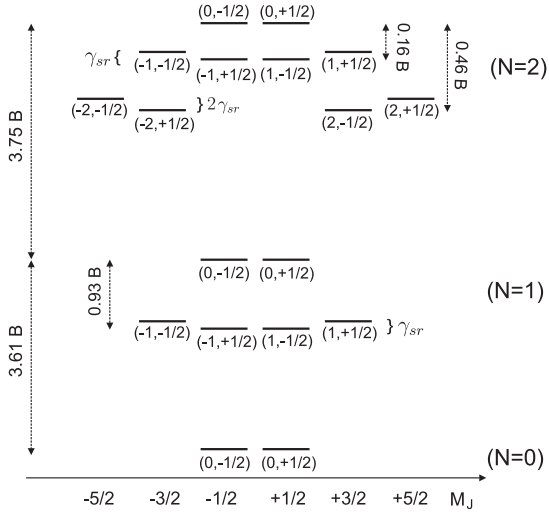}
\caption{Spectrum of a $^2\Sigma$ molecule for a bias field $E_b=3.05 B/\mu_0$. Eigenstates which
are approximately of the form $|N,M_N\rangle_{E_b}\otimes |M_S\rangle$ are individually labelled by
$(M_N,M_S)$ while the quantum number $N$ is indicated only for the corresponding manifold.}
\label{fig:SpinQubits}
\par\end{centering}
\end{figure}

In the presence of a bias field ${\bf E}_b=E_b{\bf e}_z$ the Hamiltonian for a $^2\Sigma$ molecule
in the vibrational ground state is
\begin{equation}
H_M= B{\bf N}^2 + \gamma_{sr} {\bf N}{\bf S} - \mu_z E_b\,,
\end{equation}
with $\gamma_{sr}$ the spin-rotation coupling constant, typically in the order of $100$ MHz. For a
vanishing bias field the spin-rotation coupling lifts the degeneracy of $|N,M_N\rangle_0$ states
and new eigenstates of $H_M$ are given by $|N, S; J,M_J\rangle_0$ with ${\bf J}={\bf N}+{\bf S}$
the total angular momentum. For moderate and strong electric fields $E_b\geq B/\mu_0$ the dipole
coupling typically exceeds the spin-rotation coupling and eigenstates of $H_M$ are approximately
given by product states $|N,M_N\rangle_{E_b}\otimes |M_s\rangle$. Therefore, to a good
approximation energy splitting and induced dipole moments are determined by the rotational
component only and we refer to the discussion given in Sec.~\ref{sec:RotSpectroscopy}. Especially,
for states with $M_N=0$ different spin components are degenerate and have exactly the same induced
dipole moments. For other values of $M_N\neq 0$ different spin components are split by the spin
rotation coupling, $\sim  \gamma_{sr}$, but still have to a good approximation the same induced
dipole moments. As an example we plot in Fig.~\ref{fig:SpinQubits} the spectrum of $H_M$ at the
`sweet spot' of the bias field, $E_b=E_S\equiv 3.05 B/\mu_0$.

For molecules with a finite nuclear spin $I>0$ the physical picture is quite similar although the
resulting spectrum is more involved. In the limit of $\gamma_{sr}\rightarrow 0$ the hyperfine
interaction Hamiltonian $H_{hf}$ couples the electron spin and the nuclear spin to a combined
angular momentum ${\bf F}_3={\bf S}+{\bf I}$ and eigenstates are of the form
$|N,M_N\rangle_{E_b}\otimes |F_3,M_{F_3}\rangle$. For a strong electric bias field and for $M_N=0$
this factorized form is approximately conserved even for finite $\gamma_{sr}$ as the spin-rotation
coupling is too weak to mix different rotational states. For non-zero $M_N$ the diagonal part of
the spin-rotation coupling, $\gamma_{sr}N_zS_z$, does in addition lead to a mixing of different
states $|F_3,M_{F_3}\rangle$. This general picture of the spectrum holds as long as
$|H_{hf}|,\gamma_{sr}\ll B <\mu_0E_b$ while details depend on the exact relation between hyperfine
and spin rotation coupling.


%

\subsection{Spin ensemble qubits in MDC: a protected quantum memory}

The additional spin degree of freedom allows us to encode quantum information in ensemble qubits
$|0_e\rangle$ and $|1_e\rangle$, where the corresponding molecular basis states $|g\rangle$ and
$|e\rangle$ have different spin components. In the following we explain in two examples how the
additional spin degree of freedom can compared to purely rotational states improve the lifetime of
molecular ensemble qubits in a dipolar crystal.


\subsubsection{Long-lived ensemble quantum memory}

We first consider molecular basis states
$|g\rangle\!=\!|1,0\rangle_{E_{b}}\otimes|M_S\!=\!-\!1/2\rangle$ and
$|e\rangle\!=\!|2,0\rangle_{E_{b}}\otimes|M_S\!=\!1/2\rangle$ for the bias field $E_b=E_S$. As we
discuss in Sec.~\ref{sec:Specific} this is an example for two states which can be both trapped in
electrostatic fields. The states $|g\rangle$ and $|e\rangle$ have different rotational components
and different spin components. At this point it is important to note that the factorization of
eigenstates of $H_M$ in a rotational and a spin component is only approximate. The spin-rotation
coupling still admixes a small fraction of the other spin component which results in a finite
transition dipole matrix element $|\langle e|\bs{\mu}|g\rangle|$, which actually allows us to
couple states $|g\rangle$ and $|e\rangle$ to the microwave cavity. The dimensionless transition
matrix element $\Theta_{x}=|\langle e|\mu_{x}|g\rangle|/\mu_{0}$ which determines the coupling to
the cavity as well as the parameter $\kappa$ depends on the ratio $\gamma_{sr}/B$. For the typical
case of $\gamma_{sr}\ll B$ we find in numerical calculations
$\Theta_{x}\simeq2.5\times\gamma_{sr}/B$. For molecules like CaBr with $\gamma_{sr}/B=0.03$ the
spin forbidden transition is roughly a factor 7 smaller than the corresponding spin-allowed
transition. In contrast the dimensionless parameter $\kappa \sim \Theta_x^2$ is reduced by a factor
$(\gamma_{sr}/B)^{2}$ and exciton phonon interactions are strongly suppressed.  For our specific
example we obtain $\kappa\simeq 250\times(\gamma_{sr}/B)^{2}\ll 1$, which should be compared to
$\kappa\approx 10$ for spin conserving transitions (see row a) in Table~\ref{tab:qubits}).
Therefore, the spin degree of freedom provides an additional knob to change the interaction
parameters $\kappa$ and $\epsilon$ independently and tune the crystal into the weak coupling regime
where decay rates are highly reduced (see discussion given in Sec.~\ref{sec:Lifetime}).

\subsubsection{Protected spin ensemble quantum memory}

To improve the lifetime of the ensemble quantum memory even further we consider in a next step the
states $|g\rangle\!=\!|1,0\rangle_{E_{b}}\otimes|M_S\!=\!-\!1/2\rangle$ and
$|s\rangle\!=\!|1,0\rangle_{E_{b}}\otimes|M_S\!=\!1/2\rangle$. As $|g\rangle$ and $|s\rangle$ have
the same rotational component we have introduced the new notation $|s\rangle$ to distinguish this
state from the rotationally excited state $|e\rangle$ discussed above. While the states $|g\rangle$
and $|s\rangle$ can not directly be coupled with a single microwave photon,  we can still achieve a
coupling to the microwave cavity by a two photon process involving an additional classical
microwave field.

As explained above the two spin states $|g\rangle$ and $|s\rangle$ have the same induced dipole
moment, $\epsilon=0$, and in addition there are no resonant dipolar exchange processes between
molecules in states $|g\rangle$ and $|s\rangle$ which implies that also $\kappa=0$. Therefore,
quantum information encoded in \emph{spin ensemble qubits} $|0_{s}\rangle\equiv|0_{e}\rangle$ and
$|1_{s}\rangle\equiv S_{e}^{\dag}|0_{s}\rangle=1/\sqrt{N}\sum_{i}|g_{1}\dots s_{i}\dots
g_{N}\rangle$ is naturally protected from dipole-dipole interactions and the resulting phonon
induced decay. Higher order spin flip processes due to virtual excitations into higher rotational
states~\cite{KremsSpinFlip,MicheliLattice}, which are not included in our model scale as $ \sim
\mu_0^4/(16\pi^2\epsilon_0^2 a_0^6)\times\gamma^2_{sr}/B^3$ and even for $a_{0}=50$ nm, this rate
is only in the order of a few Hz. Therefore, the lifetime of spin ensemble qubits is mainly limited
by the dephasing rate of the spin (or hyperfine states) of a single molecule. Similar to the case
of cold atoms~\cite{Cornell2002,Treutlein2004} or trapped
ions~\cite{WinelandHyperfine2005,BlattHyperfine2005} dephasing rates of molecular hyperfine states
of below 1 Hz should be achievable.

While highly protected from dipolar interactions spin ensemble qubits are not directly coupled to
microwave photons and the storage and retrieve operations require a two step process
\begin{equation}\nonumber
\left(\alpha|0_c\rangle +\beta|1_{c}\rangle\right)\leftrightarrow\left(\alpha|0_e\rangle
+\beta|1_{e}\rangle\right)\leftrightarrow\left(\alpha|0_s\rangle +\beta|1_{s}\rangle\right)\,,
\end{equation}
which involves the rotationally excited ensemble state $|1_e\rangle$. During this process the
rotational ensemble qubit is affected by interactions with phonons which cause a decay into
orthogonal states $|1_{\bf k}\rangle$ as discussed in Sec.~\ref{sec:Lifetime}. Therefore, the
overall fidelity of an ensemble quantum memory in a MDC is still affected by exciton-phonon
interactions. However, by employing a Raman process as discussed in Ref.~\cite{molensemble}
rotational states are only virtually populated and interactions with phonons are suppressed by the
detuning $\Delta=\omega_c-\omega_{eg}$.  As we have not yet analyzed the details of the
cavity-ensemble coupling we postpone the discussion of gate fidelities to Sec.~\ref{sec:Gates}
where we study swap operations between microwave photons and spin ensemble qubits for a specific
setup.

\section{Dipolar Crystals in a trap: Interfacing molecular ensembles and Circuit QED} \label{sec:Specific}
In  Sec.~\ref{sec:Rotation} and Sec.~\ref{sec:Spin} we have studied an idealized homogeneous MDC
and identified the exciton-phonon interaction as the main source of decoherence for ensemble
qubits. We have shown that for certain choices of molecular states, in particular states with a
different spin component, phonon induced decay processes are suppressed and a highly protected
ensemble quantum memory can be realized with this system. However, so far we have ignored questions
related to the experimental implementation of a MDC, especially questions of transverse and
longitudinal trapping potentials. In this section we study the properties of molecular dipolar
crystals under realistic experimental conditions, especially in the presence of an additional
longitudinal trapping potential.

For a coherent integration of a MDCs with a circuit QED system as shown in Fig.~\ref{fig:MDCSetup}
 trapping of polar molecules close above the chip surface must not affect coherence
properties of the superconducting device. In particular trapping techniques which require the
application of strong magnetic fields or intense laser fields raises compatibility questions with
high Q-values of superconducting strip line cavities~\cite{HighQ2}. Therefore, to achieve a strong
transverse confinement which is compatible with long coherence times of the microwave cavity, we
focus in this paper on a scenario where molecules are trapped by an electrostatic potential as
recently proposed in Ref.~\cite{singlemol}.
In Sec.~\ref{sec:ElectricTraps} with briefly outline the basic idea of electrostatic trapping of
polar molecules and show that this specific trap design will restrict our choice of molecular basis
states to a very limited set of rotational states. In Sec.~\ref{sec:Harmonic} we then study the
properties of excitons and phonons in a quasi 1D trapping configuration with an additional harmonic
confinement potential along the crystal axis. As the confining potential removes the translational
symmetry of the crystal, symmetric excitations $|1_e\rangle$ are no longer eigenstates of $H_{\rm
exc}$ which opens a new decay channel for rotational ensemble qubits. In Sec.~\ref{sec:Stability}
we use the spectra of longitudinal and transverse phonon modes to discuss the stability of a 1D
crystal in the case of finite temperature and a finite transverse trapping frequency. Finally, in
Sec.~\ref{sec:Gates} we use these results to discuss state transfer fidelities between a microwave
cavity and spin ensemble quibts under realistic experimental conditions.

\subsection{Electrostatic confinement of polar molecules}\label{sec:ElectricTraps}

In the following we consider a $^2\Sigma$ molecule as discussed in Sec.~\ref{sec:SpectroscopySpin},
in the presence of a bias  field ${\bf E}_b({\bf r})$ which now depends on the position of the
molecule. The Hamiltonian is
\begin{equation}\label{eq:HmolPosition}
H_M({\bf  r})= B{\bf N}^2 + \gamma_{sr} {\bf N}{\bf S} - \bs{\mu} {\bf E}_b({\bf r})\,.
\end{equation}
To achieve trapping we consider an electric field of the form ${\bf E}_b({\bf r})=(E_{\rm off} +
E_t ({\bf r})) {\bf e}_z$, with a large offset field $E_{\rm off}$ and a small trapping field
$E_t({\bf r})$ with $ 0\leq E_t({\bf r}) \ll E_{\rm off}$ and $E_t({\bf r})=0$ at the center of the
trap. Without going into the  details of the actual trap design we here envision an elongated
version of the electric z-trap trap proposed in Ref.~\cite{singlemol} which would produce an
electric field configuration of approximately this from.

As long as the position dependent trapping field $E_t({\bf r})$ is small compared to the offset
field $E_{\rm off}$ we can use a Born-Oppenheimer argument and write
Hamiltonian~\eqref{eq:HmolPosition} as
\begin{equation}
H_M({\bf r})\simeq \sum_n (E_n+ V_{t,n}({\bf r})) |n\rangle\langle n| \,,
\end{equation}
with $|n\rangle$ ($E_n$) denoting eigenstates (eigenvalues) of $H_M$ for the bias field ${\bf
E}_b=E_{\rm off}{\bf e}_z$  and $V_{t,n}({\bf r})= -\langle n|\mu_z|n\rangle E_t({\bf r})$. As the
electric field has a local minimum at the trap center only `weak field seekers' with $\mu_n=\langle
n|\mu_z|n\rangle <0$ are trapped in this potential. From the discussion in
Sec.~\ref{sec:RotSpectroscopy} we find that this restriction limits our choice of molecular basis
states to states with $M_N=0$ and moderate electric fields.

As indicated by the index $n$ the trapping potential $V_{t,n}$ depends in general on the internal
eigenstate $|n\rangle$. In the following we avoid this dependence by tuning the offset field to the
 `sweet spot', $E_{\rm off}=E_S$. We choose the two
spin states in the $N=1$ manifold, $|g\rangle\!=\!|1,0\rangle_{E_{S}}\otimes|M_S\!=\!-\!1/2\rangle$
and $|s\rangle\!=\!|1,0\rangle_{E_{S}}\otimes|M_S\!=\!+\!1/2\rangle$ as our single molecule basis
states for spin ensemble qubits. Employing a two photon process the two spin states can be coupled
to the microwave cavity via a third, rotationally excited state
$|e\rangle\!=\!|2,0\rangle_{E_{S}}\otimes|M_S\!=\!-\!1/2\rangle$. Restricted to those three basis
states we can write the  molecular Hamiltonian as
\begin{equation}
H_M({\bf r})\simeq \hbar \omega_{eg} |e\rangle\langle e| + V_{t}({\bf r})\,,
\end{equation}
with $V_{t}({\bf r})$ a state independent trapping potential for the molecule. Below we consider a
quasi 1D trapping configuration with
\begin{equation}\label{eq:Vtrap}
V_{t}({\bf r})= \frac{1}{2}m \nu x^2 + \nu_\perp^2 (y^2+z^2)\,.
\end{equation}
Here $\nu_\perp$ is the trapping frequency for the strong transversal confinement and $\nu\ll
\nu_\perp$ the trapping frequency for an additional weak confinement along the crystal axis. In
this electrostatic trap the transverse trapping frequencies can be as high as $\nu_\perp/2\pi
\approx 1-10$ MHz~\cite{singlemol}.

\subsection{MDC under quasi 1D trapping conditions}\label{sec:Harmonic}
In this section we study the properties of a MDC in a quasi 1D trapping configuration where
compared to the discussion given for a homogeneous system in Sec.~\ref{sec:Rotation} we add a
finite longitudinal and transverse trapping potential $V_{t}({\bf r})$ as given in
Eq.~\eqref{eq:Vtrap}. Restricted to the basis states $|g\rangle$, $|s\rangle$ and $|e\rangle$
identified in Sec.~\ref{sec:ElectricTraps} we extend the crystal Hamiltonian $H_{\rm MDC}$ given in
Eq.~\eqref{eq:MDC0} by the trapping potential $V_t({\bf r})$ and write the total Hamiltonian for
the inhomogeneous MDC as
\begin{equation}
H_{{\rm MDC}}=\sum_{i}\left(\frac{{\bf p}_{i}^{2}}{2m}+V_t({\bf
r}_i)+\hbar\omega_{eg}|e_{i}\rangle\langle e_{i}|\right)+\hat{V}_{dd}(\{{\bf
r}_{i}\})\,,\label{eq:MDCH}
\end{equation}
with the dipole-dipole interaction $\hat V_{dd}(\{{\bf r}_{i}\})$ given in Eq.~\eqref{eq:Vdd}. We
proceed as in Sec.~\ref{sec:Rotation} and describe dipole-dipole interaction by the induced dipole
moment $\mu_g$ and the dimensionless operator $\hat K_{ij}$. At the `sweet spot' with $\epsilon=0$
the latter is given by
\begin{equation}
\hat K_{ij}= \kappa \left(|g_i,e_j\rangle\langle e_i,g_j| +|e_i,g_j\rangle\langle
g_i,e_j|\right)\,.
\end{equation}
For our specific choice of rotational states $|g\rangle$ and $|e\rangle$ we find $\kappa\simeq
10.5$ (see Table~\ref{tab:qubits}, example a)). Note that we can omit resonant dipolar interaction
between states $|s\rangle$ and $|e\rangle$ as long as most molecules remain in state $|g\rangle$.

Assuming a crystalline phase we replace the molecular position operators by ${\bf r}_i= {\bf
r}^0_i+ {\bf x}_i$, with ${\bf r}^0_i=(x_i^0,0,0)$ and $x_i^0$ the classical equilibrium position
along the crystal axis. As molecules are confined by an additional longitudinal trapping potential
the positions $x_i^0$ are no longer equidistant. In contrast to the discussion given in
Sec.~\ref{sec:Rotation} we here also include fluctuations along the transverse directions, e.g.
${\bf x}_i=(x_i,y_i,z_i)$, to study the effect of a strong but finite transverse confinement.
Expanding $H_{\rm MDC}$ up to the lowest relevant order in ${\bf x_i}$ and decompose the crystal
Hamiltonian into an exciton part, a phonon part and exciton-phonon interactions,
\begin{equation}
H_{{\rm MDC}}=H_{{\rm exc}}+H_{{\rm phon}}+H_{{\rm int}}\,.\label{eq:Hcry1}
\end{equation}
In terms of equilibrium positions $x_i^0$ the exciton Hamiltonian is  given by
\begin{equation}\label{eq:HexcH}
H_{{\rm exc}}=\hbar \omega_{eg}\sum_{i}|e_{i}\rangle\langle
e_{i}|+\frac{\mu_g^2}{8\pi\epsilon_0}\sum_{i\neq
j}\frac{\hat{K}_{ij}}{|x^0_{i}-x^0_{j}|^{3}}\,.\end{equation} The phonon Hamiltonian contains now
both longitudinal as well as transversal phonons, $H_{{\rm phon}}=H_{{\rm
phon}}^{\parallel}+H_{{\rm phon}}^{\perp}$. Primarily we are interested in the longitudinal part
which is of the form
\begin{equation}
H_{{\rm phon}}^{\parallel}=\sum_{j}\frac{p_{x,j}^{2}}{2m}+\frac{1}{2}m\nu^2
x_{j}^{2}+\frac{6\mu_g^2}{8\pi\epsilon_0}\sum_{i\neq
j}\frac{(x_{i}-x_{j})^{2}}{|x^0_{i}-x^0_{j}|^{5}}\,.\label{eq:HphonPar}\end{equation} Although our
focus is placed on the quasi one dimensional regime with transverse motion frozen out we include
for the moment  the Hamiltonian for the transverse phonons, $H_{{\rm phon}}^{\perp}$, to study the
validity of the quasi 1D approximation. It is given by
\begin{equation}
\begin{split}H_{{\rm phon}}^{\perp}= & \sum_{j}\frac{p_{y,j}^{2}}{2m}+\frac{p_{z,j}^{2}}{2m}+\frac{1}{2}m\nu_{\perp}^{2}(y_{j}^{2}+z_{j}^{2})\\
 & -\frac{3\mu_g^2}{16\pi\epsilon_0}\sum_{i\neq j}\frac{(y_{i}-y_{j})^{2}+3(z_{i}-z_{j})^{2}}{|x^0_{i}-x^0_{j}|^{5}}\,.\end{split}
\label{eq:Hphon_perp}\end{equation} Note that for fluctuations in $z$-direction we have included
the anisotropy of the dipole-dipole interactions and obtained and additional factor of 3 (see
second line of Eq.~\eqref{eq:Hphon_perp}) which lifts the degeneracy of the two transverse phonon
branches.

The leading contribution of $H_{{\rm int}}$ is linear in the position operators $x_{i}$ and is
given by \begin{equation}\label{eq:HintH} H_{{\rm
int}}\simeq-\frac{3\mu_g^2}{8\pi\epsilon_0}\sum_{i\neq
j}\frac{x^0_{i}-x^0_{j}}{|x^0_{i}-x^0_{j}|^{5}}(x_{i}-x_{j})\otimes\hat{K}_{ij}\,.\end{equation}
Transverse phonons couple to excitons only in second order of $y_i$ and $z_i$ and can be omitted in
a first approximation.

Our goal in the following is, to study the properties of excitons and phonons in a MDC in the
presence of an additional harmonic confinement potential and compare it with the homogeneous
crystal studied in Sec.~\ref{sec:Rotation}. Since for an inhomogeneous system momentum is no longer
a good quantum number we introduce a general set of exciton eigenstates,
\begin{equation}
|1_n\rangle \equiv R_{n}^{\dag}|0_e\rangle=\sum_{i}C_{n}(i)|g_1 \dots e_i\dots
g_N\rangle\,,\label{eq:EigenOP}\end{equation} where for $ n=1\dots N$ the $C_{n}(i)$ are the
normalized coefficients of the eigenstates of $H_{\rm exc}$  and $R_n^\dag$ the corresponding set
of (approximate) bosonic creation operators. In analogy, we express longitudinal displacement
operators $x_i$ in terms of phonon annihilation (creation) operators $a_n$ ($a_n^\dag$),
\begin{equation}\label{eq:NormalModesH}
\hat x_i=\sum_{n=1}^N \sqrt{\frac{\hbar}{2m \omega(n)}}\, c_n(i)\,\left(a_n +a_n^\dag \right) \,,
\end{equation}
with $c_n(i)$ the mode function for phonons of frequency $\omega(n)$ . Ignoring transverse phonons
for the moment, the total Hamiltonian for the inhomogeneous dipolar crystal is then of the form
\begin{equation}\label{eq:MDCH2}
\begin{split}H_{{\rm MDC}} & =\sum_{n}E(n)R_{n}^{\dag}R_{n}+\sum_{m}\hbar\omega(m)a_{m}^{\dag}a_{m}\\
 & +\sum_{m,n,n'}M(m,n,n')(a_{m}+a_{m}^{\dag})\,R_{n}^{\dag}R_{n'}\,,\end{split}
\end{equation}
and we are left with the evaluation of the exciton spectrum $E(n)$, the phonon spectrum $\omega(m)$
and the coupling matrix elements $M(m,n,n')$. Although in the inhomogeneous case we cannot derive
exact analytic expressions for those quantities, good approximate solutions can be found in the
limit of large $N$. The derivations of the following results which are summarized in
App.~\ref{app:excitons} and~\ref{app:phonons} are based on a similar calculation by Morigi \emph{et
al.}~\cite{Morigi} for the phonon spectrum of a harmonically confined ion crystal.

\subsubsection{Density profile}\label{sec:density}

\begin{figure}
\begin{center}
\includegraphics[width=0.49\textwidth]{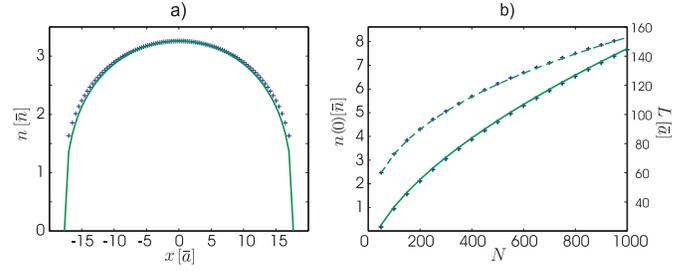}
\caption{ a) Density profile $n(x)$ for a harmonically confined dipolar crystal for $N=100$ plotted
in units of $\bar a=(m\nu^2/C_3)^{1/5}$ and $\bar n=1/\bar a$. b) Length $L$ (\emph{solid line})
 and center density $n(0)$ (\emph{dashed line}) are plotted as a function of the
particle number $N$. In both figures analytic solutions (\emph{lines}) derived in
Sec.~\ref{sec:density} are compared a numerical calculations (\emph{crosses}).} \label{fig:density}
\end{center}
\end{figure}

In a first step we need to determine the equilibrium positions of $N$ dipoles confined by the
harmonic trapping potential, $V_{t}(x)=m\nu^{2}x^{2}/2$. To do so we define the density
$n(x^0_{i})=1/|x^0_{i+1}-x^0_{i}|$ which in the limit of large $N$ and a slow variation of the
trapping potential becomes a continuous function $n(x)$. In this limit the total potential energy
of the crystal is
\begin{equation}
E_{{\rm pot}}=\int dx\left(\frac{1}{2}m\nu^{2}x^{2}-\delta_{c}\right)n(x)+\zeta(3)C_3\, n^{4}(x)\,,
\end{equation}
with $C_3=\mu_g^2/(4\pi\epsilon_0)$ and $\delta_{c}$ the chemical potential to fix the particle
number. The energy $E_{{\rm pot}}$ is minimized for the density
\begin{equation}\label{eq:density}
n(x)=n(0)\sqrt[3]{1-4x^{2}/L^{2}}\,,
\end{equation}
 where the density at the center of the trap $n(0)$ and the length
of the crystal, $L$, are given by
\begin{eqnarray}
n(0) & = & \frac{\Lambda^{2/5} }{2\zeta(3)^{1/5}} \times N^{2/5}\times \sqrt[5]{\frac{m\nu^{2}}{C_3}}\,,\label{eq:centerdensity}\\
L & = & \Lambda \times N\times a_{0}\,,\qquad a_{0}\equiv n(0)^{-1}\,.\label{eq:length}
\end{eqnarray}
Here we introduced the numerical constant $\Lambda=5\Gamma(5/6)/(\Gamma(1/3)\sqrt{\pi})\simeq
1.19$. In Fig.~\ref{fig:density} we compare the analytic result for $n(x)$ with a numerical
evaluation of the equilibrium positions $x_i^0$ for molecule numbers up to $N=1000$. We find
excellent agreement between numerical and analytic results even for a small number of molecules.

 Eqs.~\eqref{eq:centerdensity} and~\eqref{eq:length} express the
dependence of the center density and the length of the crystal as a function of the molecule number
$N$ and a given trapping frequency $\nu$. For comparison with the case of a homogeneous crystal
discussed in Sec.~\ref{sec:Rotation} it is more convenient to fix the density at the trap center,
$n(0)$, and adjust the trapping frequency $\nu$ accordingly. In the following we express quantities
in units of $a_{0}=1/n(0)$ and the corresponding energy scale $U_{dd}=\mu_g^2/(4\pi\epsilon_0
a_{0}^{3})$. The dimensionless parameter $\gamma=U_{dd}/(\hbar^2/ma_0^2)$ then gives the ratio
between potential and kinetic energy at the \emph{center} of the trap. Note that for a fixed number
of molecules we can use the relation \begin{equation}\label{eq:Nurel}
 2^{-5/2}\times \sqrt{\gamma
\Lambda^2/\zeta(3)} \times \hbar\nu=U_{dd}/N\,,\end{equation} to switch between the energy scales
of the trapping frequency and the dipole-dipole interaction.

\subsubsection{Exciton spectrum}

\begin{figure}
\begin{center}
\includegraphics[width=0.45\textwidth]{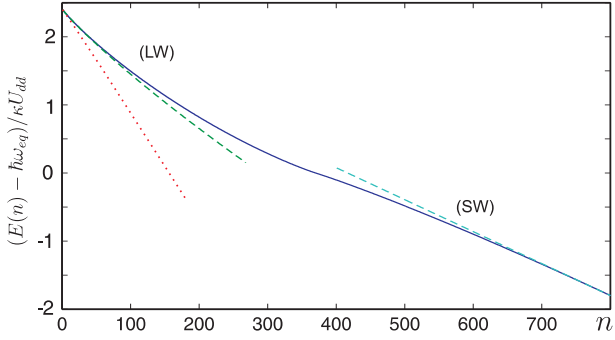}
\caption{Exciton spectrum $E(n)$ for a harmonically confined MDC for $N=800$ molecules. Numerical
results (\emph{solid line}) are compared with analytic approximations derived in
App.~\ref{app:excitons} for the long wavelength (LW) and short wavelength (SW) limit (\emph{dashed
lines}). The \emph{dotted line} indicates the LW result for the simply harmonic approximation given
in Eq.~\eqref{eq:En0} in App.~\ref{app:excitons}.} \label{fig:Excitons} \end{center}
\end{figure}

Based on set of equilibrium positions $x_i^0$ given by the density profile $n(x)$ derived above we
now evaluate the energy spectrum of the exciton Hamiltonian $H_{\rm exc}$~\eqref{eq:HexcH}. As we
show in App.~\ref{app:excitons} in the long wavelength limit and omitting the energy offset $\hbar
\omega_{eg}$ the exciton spectrum has the form
\begin{equation} \frac{E(n)}{\kappa
U_{dd}}\simeq\left[2\zeta(3)-A\sqrt{B_N+\log\left(\frac{ N}{n-1/2}\right)}\times
\frac{n-1/2}{N}\right]\,,\end{equation} with numerical constants $A=4\sqrt{\zeta(3)}/\Lambda$ and
$B_N= 3+\log(\Lambda \sqrt{\log(N/2)/32 \zeta(3)}\,)$. Similar as for a homogeneous crystal the
long range character of the dipole-dipole interactions leads to logarithmic corrections compared to
a harmonic spectrum which would result from nearest neighbor interactions. The corresponding
exciton modefunctions $C_{n}(x)$ are in a good approximation given by $C_{n}(x)\sim
\Phi_n(x,\sigma_n)$ with $\Phi_n(x,\sigma_n)$ the standard harmonic oscillator eigenfuctions
defined in Eq.~\eqref{eq:Cn0} and $\sigma_n$ a mode dependent width
\begin{equation}
\sigma^2_n\simeq N\frac{\Lambda}{4\sqrt{\zeta(3)}}\left[B_N+ \log\left( \frac{N}{n-1/2}
\right)\right]^{\frac{1}{2}}\,.
\end{equation}
In the short wavelength limit the exciton energies decrease linearly with increasing $n$ down to a
minimum energy $E_{{\rm min}}=-3\zeta(3)/2\times\kappa U_{dd}$. The corresponding, rapidly
oscillating modefunctions can be written as $C_{n}(x^0_{i})=(-1)\tilde{C}_{n}(x^0_{i})$ with
$\tilde{C}_{n}(x^0_{i})\sim \Phi_{N-n+1}(x,\sigma)$ a slowly varying envelop function of width
$\sigma\simeq 0.61\,\sqrt{N}$. Since the fast oscillations in the short wave length limit wash out
the effect of long range interactions the spectrum is purely harmonic. The analytic results are
compared with a numerical diagonalization of $H_{\rm exc}$  in Fig.~\ref{fig:Excitons}. Note that
the total the width of the exciton spectrum, $\Delta E=7\zeta(3)\kappa U_{dd}/4$ is exactly the
same as in the homogeneous case. This is due to the fact that both, long and short wavelength
excitons are located at
the center of trap where the density is almost constant. %

\subsubsection{Phonon spectrum}
\begin{figure}
\begin{center}
\includegraphics[width=0.4\textwidth]{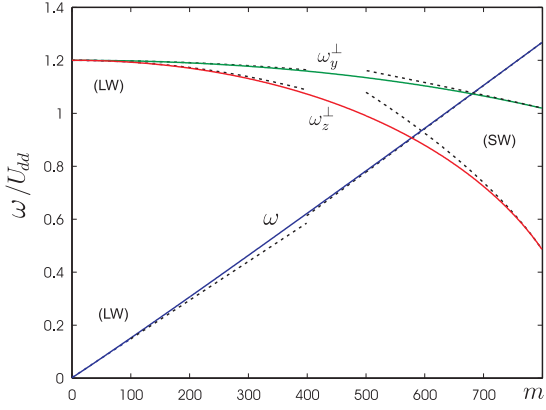}
\caption{Phonon spectrum of a harmonically confined MDC under quasi 1D trapping conditions for
$N=800$ molecules. The solid lines show the numerically evaluated spectrum for longitudinal
($\omega$) and transverse ($\omega_{y,z}^\perp$) phonons in units of $U_{dd}$ and for $\gamma=30$
and $ \hbar \nu_\perp/U_{dd}=1.2$. The dashed lines show the analytic results derived in
App.~\ref{app:phonons} for the long wavelength (LW) and short wavelength (SW) limits of the 3
different branches.} \label{fig:Phonons}
\end{center}
\end{figure}

The phonon spectrum of a quasi 1D MDC with a strong transverse trapping frequency $\nu_\perp$ and a
weak longitudinal trapping frequency $\nu$ consists of one longitudinal (`acoustic') and two
transverse (`optical') phonon branches which are plotted in Fig.~\ref{fig:Phonons}. In
App.~\ref{app:phonons} we derive approximate analytic expressions for the long and short wavelength
limits of the individual branches. For the moment we restrict our discussion to the longitudinal
modes and come back to some properties of transverse phonons in Sec.~\ref{sec:Stability}.

In the long wavelength  limit the spectrum of  longitudinal phonons determined by $H^\parallel_{\rm
phon}$ defined in Eq.~\eqref{eq:HphonPar} is of the form
\begin{equation}\label{eq:LW}
\qquad\omega(m)\simeq \nu\sqrt{1+\left(3m^{2}-m-2\right)/2}\,.
\end{equation}
 We recover the exact results for the center of mass mode $\omega(1)=\nu$ and the breathing mode
$\omega(2)=\sqrt{5}\nu$ and obtain a roughly linear phonon spectrum $\omega(m)\simeq\nu\times
1.22\times m$ for larger $m$. The corresponding  modefunctions $c_m(i)$ describe collective
oscillations extended over the whole crystal.

In the short wavelength limit the spectrum of longitudinal phonos is given by
\begin{equation}\label{eq:SW}
\omega(m)=\omega_D\,\left[1-\sqrt{\frac{40\zeta(3)}{31\zeta(5)\Lambda^2}}\,\frac{(\bar{m}+1/2)}{N}\right]\,,
\end{equation}
with $\bar{m}=N-m$ and a Debye frequency $\omega_D=\nu N\Lambda\times
\sqrt{93\zeta(5)/64\zeta(3)}$. When we reexpress the Debye frequency in units of $U_{dd}$ using
Eq.~\eqref{eq:Nurel} we find that $\omega_D$ exactly matches the Debye frequency of a homogeneous
crystal given in Eq.~\eqref{eq:wD}. This apparent coincident is based on the fact that in the short
wavelength limit phonon modefunctions are of the form $c_m(i)\approx (-1)^{i}\tilde c_{\bar m}(i)$
with an envelop function $\tilde c_{\bar m}(i)$ that is localized at the center of the trap (see
Eq.~\eqref{eq:SWsigma} in App.~\ref{app:phonons}) is therefore not sensitive to the variation of
the density at the edges of the crystal.

In summary we find that the longitudinal phonon spectrum of a harmonically confined MDC is to good
approximation linear and it is hardly affected by the long range character of dipole-dipole
interactions. The numerical results plotted in Fig.~\ref{fig:Phonons} agree well with our analytic
expressions given in Eq.~\eqref{eq:LW} and Eq.~\eqref{eq:SW} and show that for most purposes we can
simply approximate the spectrum by $\omega(m)\simeq \omega_{D}\times m/N$ where the Debye frequency
$\omega_D$ is the same as in a homogeneous crystal of  density $n=n(0)$.

\subsubsection{Exciton-phonon interaction}

The exciton-phonon interaction Hamiltonian $H_{\rm int}$ defined in Eq.~\eqref{eq:HintH} describes
transitions between excitons in mode $C_n(i)$ and $C_{n'}(i)$ by simultaneously emitting or
absorbing phonons in mode $c_m(i)$. Due to the absence of momentum conservation for a dipolar
crystal in a trap, the transition matrix element $M(m,n,n')$ for this process depends on all three
indices and has a more complicated structure as in the homogeneous case. Using the approximate
phonon spectrum $\omega(m)\simeq \omega_{D}\,m/N$ with $\omega_D$ given in Eq.~\eqref{eq:wD} we can
write transition matrix elements as
\begin{equation}
M(m,n,n')=-\frac{\kappa U_D}{\gamma^{\frac{1}{4}}} \times \mathcal{M}(m,n,n')\,,\end{equation}
where in terms of normalized equilibrium positions $\bar x_i^0=x_i^0/a_0$ the dimensionless matrix
element is
\begin{equation}\label{eq:Mnorm}
\begin{split}
&\mathcal{M}(m,n,n')=\sqrt[4]{\frac{27}{62 \zeta(5)}}\, \sqrt{\frac{N }{ m}}\,
\frac{1}{2}\sum_{i\neq j} \frac{\bar x^0_{i}-\bar x^0_j}{|\bar x^0_{i}-\bar x^0_j|^5}\times
\\&\times( c_m(i)- c_m(j))\left(C_{n'}(i)C_n(j)+C_{n'}(j)C_n(i)\right) \,.
\end{split}
\end{equation}
By replacing the coefficients $C_n(i)$ ($c_m(i)$) by the continuous modefunctions $C_n(x)$
($c_m(x)$) derived in App.~\ref{app:excitons} and~\ref{app:phonons} it is possible to study some
general properties for the matrix elements $M(m,n,n')$. For example, for long wavelength phonons
($m/N\ll 1$) we find $c_m(i)- c_m(j)\sim c'_m(x_i^0)\sim m/N$ and recover the scaling $
M(m,n,n')\sim \sqrt{m/N}$ in analogy to the homogeneous crystal. We here do not go further into
analytic details of $M(m,n,n')$ and instead we use in our calculations below numerically evaluated
values for $\mathcal{M}(m,n,n')$.

\subsection{Stability of the crystalline phase}\label{sec:Stability}

Our analysis so far has been based on the assumption that the molecules form a linear crystal with
small fluctuations around equilibrium positions. This assumption is valid in the limit of
$\gamma\gg1$, low temperatures, $k_bT \ll U_{dd}$, and strong transverse confinement $\hbar
\nu_{\perp}\gg U_{dd}$. Since in a real experiment none of these conditions is strictly satisfied
we now  study the stability of our system for finite values of $\gamma$, $T$, $\nu_{\perp}$. We
identify three processes which destabilize the dipolar crystal. First, longitudinal fluctuations of
the molecules eventually lead to a melting of the crystalline structure. For $T=0$ and a
homogeneous system this crossover has been studied numerically in Ref.~\cite{1DLozovik,1DCitro},
but we are not aware of similar studies for finite $T$ or for finite trapping potential. Second,
for a weak transverse confinement there is a regime where the linear chain is no longer the correct
ground state and molecules order in a zig-zag configuration. This so-called `zig-zag instability'
is well known for a linear chain of trapped ions where it has been analyzed
theoretically~\cite{ZigZagTh1,ZigZagTh2,Morigi} and verified experimentally~\cite{ZigZagExp}. A
third process which has been studied in Ref.~\cite{2DBuechler,MicheliConfinement} is quantum
tunnelling of molecules into regions where dipole-dipole interactions are attractive.

\subsubsection{Longitudinal stability of a dipolar crystal}

We first study the melting of a dipolar crystal due to longitudinal quantum and thermal
fluctuations of the molecules. A rigorous treatment of this problem would in principle require to
take into account the full quantum many body theory for our 1D dipolar system which is beyond the
scope of the present work. Instead we here present a much simpler calculation assuming the validity
of the phonon Hamiltonian $H^\parallel_{\rm phon}$ and determine the parameter regime where our
model is self-consistent, i.e. where fluctuations are small compared to the mean separation of the
molecules.

\begin{figure}
\begin{centering}
\includegraphics[width=0.45\textwidth]{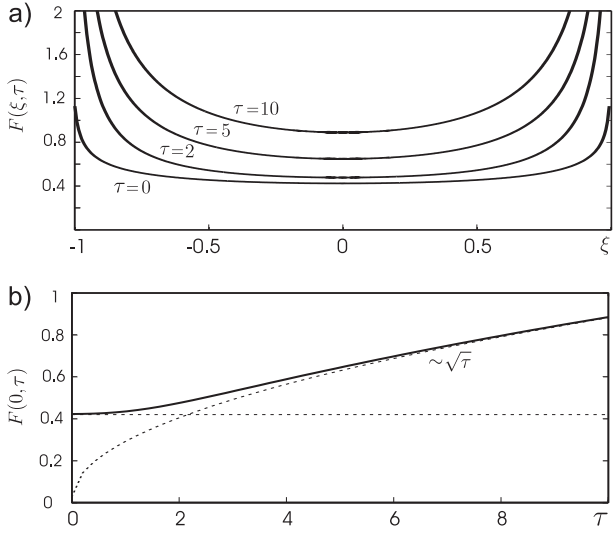}
\caption{Numerical results for the function $F(\xi,\tau)$ introduced in Eq.~\eqref{eq:Lindemann}
for $N=800$. a) $F(\xi,\tau)$ is plotted as a function of $\xi=2x/L$ for different values of
$\tau=\sqrt{\gamma}\,k_bT/U_{dd}$. b) $F_h(\tau)\equiv F(\xi=0,\tau)$ is plotted as a function of
$\tau$ (solid line). The two dashed lines indicate the two limits $F_h(\tau\ll 1)\simeq 0.42$ and
$F_h(\tau\gg1)\simeq 0.28\sqrt{\tau}$. See text for more details.} \label{fig:Fluctuations}
\par\end{centering}
\end{figure}

To study local fluctuations of the molecules in the presence of a longitudinal trapping potential
we introduce the position dependent `Lindemann parameter'
\begin{equation}\label{eq:GL} \Gamma_{L}(x,T)=n(x)\Delta_x(x,T)\,,\end{equation}
 with $(\Delta_x)^{2}(x^0_{i},T)=\langle(x_{i+1}-x_{i})^{2}\rangle$ and the average is taken with
 respect to the phonon equilibrium density operator at temperature $T$.
By employing the sound wave approximation $\omega(m)=\omega_{D}\times m/N$ we can write
$\Gamma_{L}(x,T)$ as
\begin{equation} \Gamma_{L}(x,T)=\left(\frac{1}{\gamma}\right)^{\frac{1}{4}}\times
F\left(\xi=\frac{2x}{L},\tau=\sqrt{\gamma}\,
\frac{k_{B}T}{U_{dd}}\right)\,,\label{eq:Lindemann}\end{equation} with a universal function
$F(\xi,\tau)$ defined in App.~\ref{app:fluc} in Eq.~\eqref{eq:F}.
In Fig.~\ref{fig:Fluctuations} we plot the dependence of $F(\xi,\tau)$ on temperature and position.
For the homogeneous crystal we obtain an analogous result with $F(\xi,\tau)$ in
Eq.~\eqref{eq:Lindemann} replaced by $F_h(\tau)$. Due to the nearly constant density at the center
of the trap we identify  $F_h(\tau)\equiv F(\xi=0,\tau)$. From Eq.~\eqref{eq:Lindemann} we find
that fluctuations are only weakly suppressed with increasing $\gamma$ which is already reflected in
the $\gamma^{-1/4}$ dependence of the exciton-phonon interaction (see Eq.~\eqref{eq:OHint}).

The minimal criterion for the local stability of the crystalline phase and therefore the
self-consistency of our model is $\Gamma_L(x,T)\leq 1$. To improve this criterion we argue as
follows. By Quantum Monte Carlo simulations it has been predicted~\cite{1DLozovik,1DCitro} that at
$T=0$ a crystalline phase appears for $\gamma > \gamma_c\approx 1$, or in terms of our Lindemann
parameter $\Gamma_{L}(0,0)\leq F(0,0)/\gamma_c^{1/4}\simeq 0.42$. It is reasonable to assume that a
generalization of this criterion, $\Gamma_{L}(x,T)\leq 0.42$, should also provide a good estimate
for the local existence of a crystalline phase for a finite temperature and for an inhomogeneous
density profile. Note that for the parameter regime $\gamma\approx 10-100$ this criterion
translates into $F(\xi,\tau)\leq 0.42\times \gamma^{1/4}\approx 1$. From
Fig.~\ref{fig:Fluctuations} a) we find that for $\tau=0$ fluctuations in the crystal are roughly
constant and almost the whole crystal is in a crystalline phase. For $\tau>0$ fluctuations at the
edges of the chain quickly start to increase and at around $\tau\approx 5-10$ already a significant
fraction of the system does not fulfill stability criterion. Therefore for the inhomogeneous
crystal we deduce the temperature limit $k_BT\lesssim 5 U_{dd}/\sqrt{\gamma}$. For the homogenous
crystal we find $F_h(\tau\gg1)\simeq 0.28 \sqrt{\tau}$ (see App.~\ref{app:fluc}) which leads to a
similar limit for the temperature of $k_b T\lesssim 2.4\times  U_{dd}$.

\subsubsection{Transverse stability of a dipolar crystal}

Position fluctuations of the molecules in transverse directions are described by the Hamiltonian
$H^\perp_{\rm phon}$ given in Eq.~\eqref{eq:Hphon_perp}. The resulting phonon spectrum of the two
transverse phonon branches is plotted in Fig.~\ref{fig:Phonons} and approximate analytic
expressions for the long  and short wavelength limit can be found in App~\ref{app:phonons} in
Eqs.~\ref{eq:spec_perpLW} and~\ref{eq:spec_perpSW}. As transverse phonons are hardly influenced by
the longitudinal trapping potential we use in the following discussion the exact spectrum of
transverse phonons in a homogeneous crystal, which can be written in a closed form as
\begin{equation}
\omega^{\perp}_{y,z}(q)=\sqrt{\nu_{\perp}^{2}-\frac{\alpha_{y,z}}{4\gamma}\left(\frac{U_{dd}}{\hbar}\right)^{2}\,
f^{2}(q)}\,.\label{eq:wperp}\end{equation} Here $\alpha_y=1$, $\alpha_z=3$ and $f(q)$ is given in
Eq.~\eqref{eq:f}. From Eq.~\eqref{eq:wperp} we see that transversal phonons are `optical' phonons
with a offset frequency $\nu_{\perp}$ and in the quasi 1D limit, $\hbar\nu\gg U_{dd}$ they from a
flat band, $\omega^{\perp}_{y,z}(q)\approx\nu_{\perp}$. However, when $\hbar\nu_{\perp}$ and the
dipole-dipole interaction $U_{dd}$ are of the same order we find a significant reduction of
$\omega^{\perp}_{x,y}(q)$ for short wavelength phonons (see Fig.~\ref{fig:Phonons}). In particular
the transverse phonons along z-direction reach a minimum frequency of
\begin{equation} \omega_{{\rm
min}}^{\perp}=\sqrt{\nu_{\perp}^{2}-\frac{279\zeta(5)}{8\gamma}\left(\frac{U_{dd}}{\hbar}\right)^{2}}\,.\end{equation}
 The meaning of $\omega_{{\rm min}}^{\perp}\rightarrow0$ is that the linear chain of molecules
 becomes a metastable configuration and a new ground state appears where molecules are ordered in
 a `zig-zag' configuration.  To avoid this `zig-zag' instability the minimum requirement for the transverse confinement frequency is
\begin{equation}\label{eq:ZigZag}
 \hbar\nu_{\perp}>6.08\times U_{dd}/\sqrt{\gamma}\,.
\end{equation}

While the `zig-zag' instability is a purely classical effect there is also a quantum mechanical
instability of a quasi 1D dipolar crystal due to tunnelling events into the region of attractive
dipole-dipole interactions. For a detailed description of this process the reader is referred to
Ref.~\cite{MicheliConfinement} where this tunnelling rate has been analyzed for the case of two
molecules approaching each other under 1D or 2D trapping conditions. When we adopt the result
derived in Ref.~\cite{MicheliConfinement},  Sec.~III~A~4 and use $\omega_D$ as the attempt
frequency we a obtain tunnelling rate
\begin{equation}
\Gamma_{\rm tun}\simeq \omega_D \exp\left(- c \sqrt[5]{\gamma^3 \hbar \nu_\perp/(8
U_{dd})}\right)\,,
\end{equation}
with a numerical constant $c\approx 5.8$. When we assume that the inequality~\eqref{eq:ZigZag} is
satisfied we find that the tunnelling rate is bound from above by $\Gamma_{\rm tun} \lesssim
\omega_D \exp(-c\sqrt{\gamma})$. We conclude that when the system is in a crystalline regime,
$\gamma >1$, and condition~\eqref{eq:ZigZag} is fulfilled the crystal is also stable with respect
to  tunnelling events.

\subsubsection{Summary}
In summary we find that apart from $\gamma > 1$ the stability of a homogeneous quasi 1D dipolar
crystal requires the following relation between the relevant energy scales,
\begin{equation}\label{eq:ineq}
{\rm max}\left[\frac{\hbar^2}{m a_0^2},0.42\, k_{B}T \right] < U_{dd}< \sqrt{\gamma}\,
\frac{\hbar\nu_{\perp}}{6.08}\,.\end{equation} In the inhomogeneous case the bound on the
temperature should be replaced by $k_BT\leq 5\, U_{dd}/\sqrt{\gamma}$ to guarantee a crystalline
structure over a large fraction of the system. Finally we emphasis that numerical values for the
bound on the temperature should not be considered as precise numbers. One hand this is due to our
simplified model we used to calculate these numbers. On the other hand in a 1D configuration we
anyway expect a smooth crossover from a crystalline to a liquid regime with no precisely defined
transition point. In that sense Eq.~\eqref{eq:ineq} determines the parameter regime where our model
Hamiltonian $H_{\rm MDC}$~\eqref{eq:MDCH2} is valid. Outside this regime we do not expect a
complete break down of our model but higher order corrections should be taken into account.

\subsection{State transfer fidelities for spin ensemble qubits}\label{sec:Gates}

So far in Sec.~\ref{sec:Specific} we have discussed several aspects of a MDC in the presence of a
finite longitudinal and transverse confinement potential and we have identified the stability
criteria for such a system. In particular we have focused on a specific scenario where molecules
are trapped in electrostatic potentials which provides a convenient way to achieve strong
transverse confinement and the possibility to bring molecules close the surface of the cavity
electrode to enhance the coupling strength. However, the severe restriction on trappable rotational
states limits the induced dipole moments to $\mu_g\approx 0.2 \mu_0\approx 1D $ while at the same
time resonant dipole-dipole interaction and therefore exciton-phonon interactions are quite large
($\kappa\approx 10$). In this section we finally want to show that even under those unfavorable
conditions we may still achieve  high fidelity state transfer operations between the cavity and
ensemble qubits  encoded in collective spin excitations.

\subsubsection{Ensemble-cavity coupling}

In order to couple the two spin states $|g\rangle$ and $|s\rangle$ defined in
Sec.~\ref{sec:ElectricTraps} we consider a Raman type setup where $|g\rangle$ and the rotational
state $|e\rangle$ are coupled to the cavity field and $|e\rangle$ is in turn coupled to state
$|s\rangle$ by a classical microwave field of orthogonal polarization. For this configuration the
cavity-molecule interaction is
\begin{equation}\label{eq:Raman1}
H_{\rm cav-mol}(t)=  \sum_i g(x) |g_i\rangle\langle e_i| c^\dag +
\frac{\Omega(t)}{2}|s_i\rangle\langle e_i|+ {\rm H.c.}\,,
\end{equation}
with $\Omega(t)$ the controllable Rabi frequency of the external microwave field. In
Eq.~\eqref{eq:Raman1} we have generalize our model to a non-uniform cavity modefunction by
introducing the position dependent single molecule coupling $g(x)=g u(x)$. Here $g$ is the maximum
coupling constant at the center of the trap and $0\leq u(x)\leq 1$. Note that $u(x)$ varies at
least on the scale of the cavity wavelength $\lambda_c$, but may in principle be designed to have
an arbitrary shape. Under the two-photon resonance condition and a detuning
$\Delta=\omega_c-\omega_{eg}\gg g\sqrt{N},\Omega(t)$ we can eliminate the excited state and obtain
\begin{equation}\label{eq:Raman2}
H_{\rm cav-mol}(t)= g_{R}(t)\left(S_{e}^{\dag}c+c^{\dag}S_{e}\right)\,,
\end{equation}
with a Raman coupling strength $g_{R}(t)=g_N \Omega(t)/2\Delta$. Note that due to the non-uniform
cavity-molecule coupling we obtain a modified ensemble coupling $g_N\equiv g \sqrt{N_{\rm eff}}$
with an effective number of molecules $N_{\rm eff} =\int |u(x)|^{2}n(x)dx$. Similarly, ensemble
qubit states are defined as
\begin{equation}
|1_s\rangle\equiv S_e^\dag|0_s\rangle= \frac{1}{\sqrt{N_{\rm eff}}}\sum_i u(x_i^0)|g_1\dots s_i
\dots g_n\rangle\,.
\end{equation}

Assuming for simplicity $\Omega(t)/2\approx g_N$ the Raman coupling Hamiltonian~\eqref{eq:Raman2}
provides the basic ingredient for a swap operation between the state of the cavity and the spin
ensemble qubit in a time  $T_S=\pi \Delta /2g^2_N \sim 1/g_N$. For a realistic estimate of $g_N$ we
consider the predictions for the single molecule coupling strength $g$ which are given in
Ref.~\cite{singlemol} for CaBr with $\mu_0=4.3$ D. With $d$ the distance between the molecules and
the cavity electrode we obtain
\begin{equation}
g_N/(2\pi) \simeq  40\, {\rm kHz} \times  \sqrt{N_{\rm eff}} / d[\mu m]\,.
\end{equation}
As in the crystalline phase we do not have to care about motional diffusion of molecules the length
of the crystal can in principle be as large as $L\approx \lambda_c/2$ meaning that even for a 1D
crystal the number of molecules can be as high as $N_{\rm eff}\approx \lambda_c/(2 \Lambda
a_0)\approx 10^5$. Using a moderate trap distance $d=0.5\, \mu m$ we end up with a collective
coupling strength in the order of $g_N/(2\pi) \approx 25\,{\rm MHz}$, which can in principle be
pushed into the 100 MHz regime by going to trap-surface distances of $d\approx 0.1 \,\mu m$.

\subsubsection{Decay of rotational ensemble qubits}

As decoherence processes for spin ensemble qubits during gate operations are due to a finite
population of the rotationally excited state $|e\rangle$ we study in a first step the decay of a
rotational ensemble qubit state $|1_e\rangle$. Following the calculations of
Sec.~\ref{sec:Lifetime} the initial decay of the excited state probability is $P_e(t)\approx 1- W^2
t^2$, where compared to the homogeneous case the decay rate $W$ now contains two contributions,
\begin{equation}\label{eq:W2H}
 W^2=\left(\frac{\kappa U_{dd}}{\hbar}\right)^2\left(\mathcal{I}_{\rm exc}+\frac{\mathcal{I}_{\rm
phon}}{\sqrt{\gamma}}\right)\,.
\end{equation}
 The first term in Eq.~\eqref{eq:W2H} is the exciton dispersion which arises in an inhomogeneous system
 from the
mismatch between cavity modefunction and exciton eigenfunctions. In terms of the normalized exciton
spectrum $\bar E(n)=(E(n)-\hbar \omega_{eg})/(\kappa U_{dd})$ and the overlap $z_n=\sum_i C_n(i)
u(x_i^0)$ it is given by
\begin{equation}
\mathcal{I}_{\rm exc}=\sum_{n} \bar E^2(n)  z^2_n - \Big(\sum_n \bar E(n) z^2_n\Big)^2\,.
\end{equation}
For a cavity modefunction $u(x)=\cos(\pi x/\lambda_c)$ and $L\leq \lambda_c/2$ we obtain numerical
values of $\mathcal{I}_{\rm exc}\simeq 0.40-0.11$, where the lower value corresponds to
$L=\lambda_c/2$. Of course, a more sophisticated trap or cavity design would reduce this value even
further towards $\mathcal{I}_{\rm exc}\simeq 0$, e.g., in the limit of a flat bottom trap.

The second term in Eq.~\eqref{eq:W2H} describes the decay due to exciton-phonon interactions and is
defined as
\begin{equation}
\mathcal{I}_{\rm phon}=\sum_{m,n,n'} z_n z_{n'} |\mathcal{M}(m,n,n')|^2\left(2
N(\omega(m))+1\right)\,,
\end{equation}
with normalized coupling matrix elements $\mathcal{M}(m,n,n')$ defined in Eq.~\eqref{eq:Mnorm}. A
numerical evaluation of $\mathcal{I}_{\rm phon}$ for the case $L\ll \lambda_c$ shows that in the
zero temperature limit $\mathcal{I}_{\rm phon}\approx 1.38$ while for high temperatures we obtain
$\mathcal{I}_{\rm phon}\approx 11.3\times \tau$, with $\tau=\sqrt{\gamma}\,k_BT/U_{dd}$. The
crossover point between the two regimes is $\tau\approx 1$.

\subsubsection{State transfer fidelities for spin ensemble qubits}

For a simplified discussion of the state transfer fidelity between the microwave cavity and spin
ensemble qubits we identify the gate fidelity $\mathcal{F}$ with the probability to convert a
single cavity photon, $|1_c\rangle$, into a spin excitation $|1_s\rangle$. The fidelity is degraded
by two processes. First, the spin state $|1_s\rangle$ decays due to the finite admixture of the
rotational ensemble state $|1_e\rangle$. For a detuning $\Delta\gg |H_{\rm MDC}|$ and the fast gate
times $T_G$ considered in this paper this decay is quadratic and the corresponding rate $W_s\approx
(g_N/\Delta)^2 \times W$ is proportional to $W$ defined in Eq.~\eqref{eq:W2H}, but suppressed by
$(g_N/\Delta)^2$. Second, the photon state $|1_c\rangle$ decays linearly with
 the cavity decay rate $\Gamma_c$. Assuming that each of the two processes acts for approximately
half of the gate time $T_G$ we obtain a total gate fidelity
\begin{equation}
\mathcal{F}\simeq 1-\left(\frac{\pi W}{4\Delta}\right)^{2} -\frac{\pi \Gamma_c \Delta}{4g_N^2} \,.
\end{equation}
We optimize the gate fidelity for a detuning $\Delta_*=(\pi g_N^2W^2/\Gamma_c)^{1/3}$  which
results in a  maximal fidelity of
\begin{equation}\label{eq:FidOp}
\mathcal{F}_*\simeq 1-\frac{3}{4} \left(\frac{\pi}{2}\right)^{\frac{4}{3}}\,  \left( \frac{\Gamma_c
W}{g_N^2}\right)^{\frac{2}{3}} \,.\end{equation}

\emph{Discussion.} We now consider a specific example using the molecule CaBr with $\mu_0=4.3$ D,
$m=120$ amu and $B/2\pi= 2.8$ GHz. From  Table~\ref{tab:qubits} a) we find that at the sweet spot
the induced dipole moment is $\mu_g\approx 0.7$ D and $\kappa=10.5$. To achieve a stable crystal we
choose a lattice spacing of $a_{0}=70$ nm which corresponds to $\gamma \approx 13$ and
$U_{dd}/(2\pi)=215$ kHz. From our estimates on the stability of the crystal in
Sec.~\ref{sec:Stability} we obtain the conditions $k_BT\lesssim 2 U_{dd}\approx  20\,\mu$K and
$\nu_\perp/2\pi > 360$ kHz.  Both requirements seem feasible with on-chip cooling and trapping
techniques proposed in Ref.~\cite{singlemol}. Trapping a moderate number of molecules $N\approx
10^4$ at a distance $d\approx 0.5\,\mu$m above the cavity electrode we obtain a collective coupling
strength of $g_{N}/2\pi\approx 8$ MHz. Using a superconducting microwave cavity with a quality
factor $Q\approx 10^6$ as demonstrated in Ref.~\cite{HighQ2} the decay rate of the cavity is as low
as $\Gamma_c=\omega_c/Q\approx 2\pi\times 10$ kHz. Without any mode matching but assuming
temperatures of $k_BT\leq U_{dd}/\sqrt{\gamma}\approx 3 \,\mu$K we obtain  $W/(2\pi) \approx 2$
MHz. Inserting these values into Eq.~\eqref{eq:FidOp} we obtain a gate fidelity of $F\simeq 0.994$
and a gate time of $T_G\approx 0.14\,\mu$s. For a more optimistic choice of parameters with
$g_N/2\pi=25$ MHz and $a_0=100$ nm (which corresponds to $\gamma\approx 9$, $U_{dd}/2\pi\approx 75
$ kHz and a temperature requirement $k_B T< 1\,\mu$K) we immediately obtain gate errors of below
$10^{-3}$ at even shorter gate times.

In conclusion we find that while in an electrostatic trapping configuration exciton-phonon
interactions are quite high, it can be overcome by the high collective coupling strength due to the
high densities in the crystalline phase and the low trap surface distance. However, for further
improvements it might be necessary to consider magnetic trapping techniques where exciton-phonon
interactions can be highly reduced while at the same time temperature requirements would be less
stringent. A second interesting alternative is the choice of rotational states $|g\rangle$ and
$|e\rangle$ which are given in Table~\ref{tab:qubits}, example e). Both states are weak field
seekers \emph{and} at the same time they satisfy the decoupling condition $\kappa+\epsilon=0$.
However, as molecules in state $|g\rangle$ and $|e\rangle$ have a different induced dipole moment
and feel a different trapping potential, this configuration would require a flat bottom trap in
longitudinal direction and ground state cooling in transverse directions.

\section{Summary \& Conclusion}\label{sec:Summary}

In this paper we have investigated the storage of quantum information encoded in  collective
excitations (ensemble qubits) of long-lived rotational or spin degrees of freedom in a
self-assembled dipolar crystal of polar molecules. This provides a high fidelity quantum memory
which can be coupled to a superconducting strip line cavity which in the spirit of Cavity QED
provides a coupling to a solid state quantum processor. The main results are summarized as follows.

In the first part of this work  we have studied the dynamics of rotational excitations  (=
excitons) in a self-assembled molecular dipolar crystal (MDC) which maps to a polaron type model
with excitons interacting with the phonon modes of the crystal. While in general the exciton-phonon
interactions plays the dominant role as a decoherence mechanism in this system, leading to a decay
of rotational ensemble qubits, we have identified certain `magic' configurations where long
wavelength excitons -- which includes the ensemble qubit state -- decouple from the phonon modes.
Furthermore, quantum information encoded in spin ensemble qubits is naturally protected from
dipole-dipole interactions, and  the exciton-phonon interactions affects  the ensemble quantum
memory  only during gate operations.

 In the second part of this paper we have studied in detail a specific scenario with molecules
 trapped in electrostatic potentials (e.g. an on-chip electric trap) and quantum information encoded
 in collective spin excitations. We have discussed modification of the exciton and phonon spectrum due
 to the presence of a longitudinal trapping potential and analyzed the stability of a MDC in a quasi-1D geometry.
 An estimate of the expected state transfer fidelities between the microwave cavity and spin ensemble qubit for this
setup shows that under reasonable experimental conditions fidelities  of $\mathcal{F}\geq0.99$ can
be achieved for a total gate time well below 1 $\mu s$. Optimized conditions would result in gate
errors of $\sim 10^{-4}$ which would allow fault tolerant quantum computing~\cite{Faulttolerant}.
This specific example demonstrates the potential of MDCs in the context of hybrid quantum computing
since high gate fidelities and long storage times are combined with gate times that are compatible
with decoherence times scales in solid state based quantum computing.


\begin{acknowledgments}
We thank  H. P. B\"uchler, A. Micheli, G. Pupillo and M. Lukin for stimulating discussions and P.
Xue for contributions at the initial stage of this work. This work was supported by the Austrian
Science Foundation (FWF), the European Union projects EuroSQIP (IST-3-015708-IP), CONQUEST
(MRTN-CT-2003-505089) and SCALA (IST-15714), and the Institute for Quantum Information.
\end{acknowledgments}



\appendix

\section{Homogeneous Dipolar Crystal}
\label{app:ring} In this Appendix we briefly summarize the derivation of the exciton spectrum
$E({\bf k})$, the phonon spectrum $\omega_\lambda({\bf k})$ and the coupling matrix elements
$M_\lambda({\bf q},{\bf k})$ for a homogeneous dipolar crystal in 1D and 2D. All results in this
appendix are expressed in units of the lattice spacing $a_0$ and the corresponding dipole-dipole
energy $U_{dd}=\mu_g^2/(4\pi\epsilon_0a_{0}^{3})$. In these units the parameter
$\gamma=U_{dd}/(\hbar^2/ma_0^2)$ plays the role of the dimensionless mass of the molecules.

\emph{Excitons.} We start with the exciton Hamiltonian $H_{\rm exc}$ as defined in
Eq.~\eqref{eq:Hexc}. In dimensionless units it is given by
\begin{equation}
H_{\rm exc}=\sum_i \tilde E_{eg} |e_i\rangle\langle e_i|+ \frac{1}{2}\sum_{i\neq j}\frac{\hat
K_{ij}}{|{\bf r}^0_i -{\bf r}^0_j|^3}\,,
\end{equation}
with $\tilde E_{eg}=\hbar\omega_{eg}/U_{dd}$ and $\hat K_{ij}$ defined in Eq.~\eqref{eq:Kij}. In
the limit of a low number of rotational excitations we can expressing $H_{\rm exc}$ in terms of the
exciton operators $R^\dag_{\bf k}=1/\sqrt{N}\sum_i e^{i{\bf kr}^0_i} |e_i\rangle\langle g_i|$  by
making the substitutions
\begin{equation}\label{eq:Kijsub}
\begin{split}
\hat K_{ij}\simeq&\frac{1}{N}\sum_{{\bf k},{\bf k}'} R^\dag_{\bf k} R_{{\bf k}'}\Big[\epsilon(e^{-i
({\bf k}-{\bf k}') {\bf r}_i^0}+e^{-i ({\bf k}-{\bf k}') {\bf r}_j^0})\\&+\kappa(e^{+i {\bf k}'
{\bf r}_i^0}e^{-i {\bf k} {\bf r}_j^0}+e^{+i {\bf k}' {\bf r}_j^0}e^{-i {\bf k} {\bf
r}_i^0})\Big]\,,
\end{split}
\end{equation}
and $\sum_i |e_i\rangle\langle e_i|\simeq \sum_{\bf k} R_{\bf k}^\dag R_{\bf k}$. Evaluating the
resulting expressions we end up with a diagonal Hamiltonian of the form $H_{\rm exc}=\sum_{\bf k}
E({\bf k})R_{\bf k}^\dag R_{\bf k}$, where the energy spectrum $ E({\bf k})= \tilde E_{eg}
+\epsilon J(0) + \kappa J({\bf k})$ is given in terms of the dimensionless function
\begin{equation}\label{eq:Jfull}
J({\bf k})= \sum_{i\neq0} \frac{\cos({\bf k} {\bf r}^0_i)}{|{\bf r}^0_i|^3}\,.
\end{equation}
For the 1D crystal with $r_i=i$ we can evaluate $J(k)$ and obtain
\begin{equation}\label{eq:J1D}
J(k)=2\sum_{j=1}^{\infty}\frac{\cos(kj)}{|j|^{3}}\,= {\rm Li}_3(e^{-ik})+{\rm Li}_3(e^{ik})\,,
\end{equation}
with ${\rm Li}_n(z)$ the polylogarithm function. In 2D the equilibrium positions ${\bf r}_i^0$ form
a triangular lattice with basic lattice vectors $a_1=(1,0)$ and $a_2=(1,\sqrt{3})/2$ and we
evaluate the function $J({\bf k})$ numerically (see Fig.~\ref{fig:2DCrystal}). By replacing the
summation in Eq.~\eqref{eq:Jfull} by an integral we obtain the linear behavior, $J(|{\bf
k}|\rightarrow 0 )-J(0)\sim |{\bf k}|$, for the long wavelength limit. For a more detailed study of
the spectrum of (rotational) excitons in 2D analytical tools developed in the field of 2D Wigner
crystals~\cite{2DWigner} or 2D spin waves~\cite{2DSpinWaves} can be applied to handle the slowly
convergent sums in Eq.~\eqref{eq:Jfull}. This analysis will be the subject of future work.

\emph{Phonons.} In a next step we consider the Hamiltonian of longitudinal phonons, $H_{\rm phon}$,
which is given by
\begin{equation}
H_{\rm phon}=\sum_i \frac{{\bf p}_i^2}{2\gamma} + \frac{3}{4}\sum_{i\neq j} \frac{ 5\left[ ({\bf
x}_i-{\bf x}_j)\cdot {\bf n}^0_{ij}\right]^2 -\left[{\bf x}_i-{\bf x}_j\right]^2}{|{\bf r}_i^0-{\bf
r}_j^0|^5}\,.
\end{equation}
As $H_{\rm phon}$ is quadratic in position and momentum operators we can rewrite it terms of phonon
annihilation and creation operators, $H_{\rm phon}=\sum_{{\bf q},\lambda}\hbar \omega_\lambda ({\bf
q}) a_\lambda^\dag({\bf q})a_\lambda({\bf q})$. To find the phonon spectrum $\omega_\lambda({\bf
q})$ we change into the Heisenberg picture and make the ansatz
\begin{widetext}
\begin{equation}\label{eq:xiAnsatz}
{\bf x}_i(t) =\frac{1}{\sqrt{N}}\sum_{\bf q}\sum_{\lambda=1}^d \sqrt{\frac{1}{2\gamma
\omega_{\lambda}({\bf q})} } \,{\bf e}_\lambda({\bf q}) \, \left(a_{\lambda}({\bf q}) e^{i({\bf
q}{\bf r}^0_i- \omega_\lambda ({\bf q})t)} +a^\dag_{\lambda}({\bf q})e^{-i({\bf q}{\bf r}^0_i-
\omega_\lambda ({\bf q})t)} \right)\,,
\end{equation}
\end{widetext}
where in the 2D case the vectors ${\bf e}_\lambda({\bf q})$ are the two orthonormal polarization
vectors of the two phonon branches. This ansatz in combination with the Heisenberg equations $
\ddot {\bf x}_i(t)=-[H_{\rm phon},{\bf p}_i]/\gamma$ leads to eigenvalue equation
\begin{equation}
-\omega^2_\lambda({\bf q}){\bf e}_\lambda({\bf q})= \mathcal{A}({\bf q}) {\bf e}_\lambda({\bf
q})\,.
\end{equation}
Here $\mathcal{A}({\bf q})$ is a single valued function in 1D and a $2\times 2$ matrix in 2D. For
the 1D case we rewrite it as $\mathcal{A}(q)= f^2(q)/\gamma $ with
\begin{eqnarray}\label{eq:f}
 f^2(q)&=&48\sum_{j=1}^{\infty}\frac{\sin^{2}(qj/2)}{j^{5}}\\
 &=& 12\left[2\zeta(5) -{\rm Li}_5(e^{iq})-{\rm Li}_5(e^{-iq})\right]\,. \nonumber
 \end{eqnarray}
For the 2D crystal the matrix $\mathcal{A}({\bf q})$ is defined as
\begin{equation}
\mathcal{A}({\bf q}) = \frac{3}{\gamma}\sum_{j\neq 0}\left[1-\frac{5}{|{\bf r}^0_j|^{2}}
\left(\begin{array}{cc} (x_j^0)^2 & x^0_j y^0_j\\ x^0_j y^0_j & (y_j^0)^2
\end{array}\right)\right]\,  \frac{(1- e^{i{\bf  q} {\bf  r}^0_j})}{|{\bf r}^0_j|^{5}}\,.
\end{equation}
By numerically solving the eigenvalue problem for the matrix $\mathcal{A}({\bf q})$ for each value
of ${\bf q}$ we obtain the two phonon branches $\omega_\lambda({\bf q})$ and the corresponding
polarization vectors ${\bf e}_\lambda$. In analogy to the 1D case we express the resulting phonon
spectrum in terms of the two rescaled functions $f_\lambda({\bf q})=\omega_\lambda({\bf
q})\sqrt{\gamma}$ (see Fig.~\ref{fig:2DCrystal}).

\emph{Exciton-phonon interactions.} Finally, we consider the first order exciton-phonon
interaction, $H_{\rm int}$, given by
\begin{equation}
H_{{\rm int}}=-\frac{3}{2}\sum_{i\neq j}\frac{{\bf r}_{i}^{0}-{\bf r}_{j}^{0}}{|{\bf
r}_{i}^{0}-{\bf r}_{j}^{0}|^{5}}\,({\bf x}_{i}-{\bf x}_{j})\otimes\hat{K}_{ij}\,.
\end{equation}
Using Eqs.~\eqref{eq:Kijsub} and \eqref{eq:xiAnsatz} we reexpress the interaction Hamiltonian in
terms of exciton and phonon operators we leads us to the form of $H_{\rm int}$ given in
Eq.~\eqref{eq:Hint} with a coupling matrix element
\begin{equation}
\begin{split}
M_\lambda({\bf q},{\bf k}) =& -\frac{3}{2}\sqrt{\frac{1}{2N\gamma \omega_\lambda({\bf
q})}}\sum_{j\neq 0 } \frac{{\bf r}^0_j\cdot {\bf e}_\lambda({\bf q})}{|{\bf
r}^0_j|^5}\left(1-e^{i{\bf q r}^0_j}\right) \\&\times \left[ \epsilon \left( 1+e^{i{\bf qr}^0_j}
\right) + \kappa \left(e^{-i({\bf k} +{\bf q}){\bf  r}^0_j}+e^{i{\bf k  r}^0_j} \right)\right]\,.
\end{split}
\end{equation}
By introducing the function
\begin{equation}\label{eq:gfull}
g_\lambda({\bf q})=\frac{3}{\sqrt{2}}\sum_{j\neq 0 } \frac{{\bf r}^0_j\cdot {\bf e}_\lambda({\bf
q})}{|{\bf r}^0_j|^5}\, \sin\left({\bf q} {\bf r}^0_j\right)\,,
\end{equation}
we can bring $M_\lambda({\bf q},{\bf k})$ into the form given in
Eq.~\eqref{eq:CouplingMatrixElements}.  For 1D we can evaluate the sum analytically,
\begin{equation}\label{eq:g1D}
g(q)=\frac{6}{\sqrt{2}}\sum_{j=1}^{\infty}\frac{\sin(qj)}{j^{4}}= \frac{3}{\sqrt{2}}\left( {\rm
Li}_4(e^{-iq})-{\rm Li}_4(e^{+iq})\right)\,.\end{equation} In 2D the evaluation of the functions
$g_\lambda({\bf q})$ requires a numerical summation of the right hand side of Eq.~\eqref{eq:gfull}.

\section{Exciton Spectrum of a Harmonically Confined MDC}\label{app:excitons}
In this appendix we derive approximate analytic  expressions for the exciton spectrum in a
harmonically confined dipolar crystal in 1D. All quantities in this appendix are expressed in units
of the mean molecule separation at the center of the trap $a(0)=1/n(0)$ and the corresponding
dipole-dipole energy $U_{dd}$ (see Sec.~\ref{sec:density}).

We start with the exciton Hamiltonian $H_{\rm exc}$ given in Sec.~\ref{sec:Harmonic} in
Eq.~\eqref{eq:HexcH} and look at the eigenvalue equation $E(n)R_n^\dag=[H_{\rm exc},R_n^\dag]$ for
exciton operators $R_n^\dag = \sum_i C_n(i)|e_i\rangle\langle g_i|$ with $n=1\dots N$. In the limit
of low number rotational excitions the resulting eigenvalue equation is of the form
\begin{equation}\label{eq:ExcEig}
(E(n)-\tilde E_{eg})C_n(i)=\kappa  \sum_{j\neq i}\frac{C_n(j)}{|x_{i}^0-x_j^0|^3}\,,
\end{equation}
with $\tilde E_{eg}=\hbar \omega_{eg}/U_{dd}$. For convenience we omit this constant energy offset
in the following calculations. Our goal is to derive approximate analytic solutions of
Eq.~\eqref{eq:ExcEig} in the long and short wavelength limit.

\emph{Long wavelength limit.} For a large number of molecules $N$ the density $n(x)$ varies slowly
over the extension of the crystal and we  make the approximation
\begin{equation}\label{eq:ExcEig2}
\sum_{j\neq i}\frac{1}{|x_{i}^0-x_j^0|^3} \simeq 2\zeta(3)n^3(x^0_i)\,,
\end{equation}
to convert Eq.~\eqref{eq:ExcEig} into the form
\begin{equation}\label{eq:ExcEig3}
\left[ E(n)- 2\zeta(3)\kappa n^3(x^0_i)\right]C_n(i)= \kappa \sum_{j\neq
i}\frac{C_n(j)-C_n(i)}{|x^0_j-x^0_i|^3}.
\end{equation}
In the long wavelength limit $C_n(x_i^0)\equiv C_n(i)$ can be approximated by a slowly varying
continuous function $C_n(x)$ and
\begin{equation}\label{eq:Cexpansion}
C_n(j)-C_n(i)\simeq C_n'(x^0_i)(x^0_j-x^0_i)+C_n''(x^0_i)(x^0_j-x^0_i)^2/2 \,.
\end{equation}
Inserting this expansion into the right hand side of Eq.~\eqref{eq:ExcEig3} we obtain two
contributions. The first one proportional to the first derivative $C_n'(x)$ vanishes due to the
summation over an equal number of terms with positive and negative sign. In contrast, the second
contribution proportional $C_n''(x)$ diverges as $\sim\log(N)$. This divergence is a consequence of
the long range character of dipole-dipole interactions and means that Eq.~\eqref{eq:ExcEig3} is
sensitive to the spatial extension of the wavefunction $C_n(x)$ and the expansion in
Eq.~\eqref{eq:Cexpansion} will fail to predict accurate results. To handle this difficulty we
proceed as follows. In a first step we approximate the divergent sum by
\begin{equation}\label{eq:divsum}
\sum_{j\neq i}\frac{1}{|x^0_{i}-x^0_j|} = 2\Sigma_0 n(x_i^0) \,,
\end{equation}
where in a zeroth order approximation we set $\Sigma_0=\log(N/2)$. Under this approximation we find
that (see below) wavefunctions $C_n(x)$ are harmonic oscillator eigenfunctions. In a second step we
use the spatial dependence of these zeroth order eigenfunctions  for a more accurate reevaluation
of the right hand side of Eq.~\eqref{eq:ExcEig3}.

Using the ansatz in Eq.~\eqref{eq:divsum} we can convert the eigenvalue equation~\eqref{eq:ExcEig3}
into the differential equation,
\begin{equation}
\left[E(n)- 2\zeta(3)\kappa n^3(x)\right]C_n(x)= \kappa  \Sigma_0 n(x) C_n''(x) \,.
\end{equation}
For the density profile $n(x)$ given in Eq.~\eqref{eq:density} and with a  rescaled variable
$y=2x/L$ we obtain
\begin{equation}\label{eq:Diffeq1}
 C_n''(y)+ b^2\left[\frac{\alpha_n -y^2}{(1-y^2)^{1/3}}\right] C_n(y)=0 \,,
\end{equation}
with $\alpha_n=1-E(n)/(2\zeta(3)\kappa)$ and $b^2=\zeta(3)L^2/(2\Sigma_0)$. In a final
approximation we expand Eq.~\eqref{eq:Diffeq1} to lowest order in $y^2$ and $\epsilon_n$ and end up
with the differential equation
\begin{equation}
 C_n''(y)+ b^2(\alpha_n-y^2) C_n(y)=0 \,.
\end{equation}
Solutions of this equation are $C_n(x)\sim\Phi_{n}(x,\bar \sigma)$ with $\bar \sigma^2=
L\sqrt{\Sigma_0/8\zeta(3)}$ and $\Phi_n(x,\bar \sigma)$ the well-known harmonic oscillator
functions
\begin{equation}\label{eq:Cn0}
\Phi_n(x,\sigma)= H_{n-1}(x/ \sigma)\,e^{-\frac{x^2}{2 \sigma^2}}\,,
\end{equation}
with  $H_n(x)$ Hermite polynomials.  Using $L=\Lambda\,N$ the corresponding harmonic energy
spectrum is
\begin{equation}\label{eq:En0}
 E(n)=\kappa \left[2\zeta(3)-\frac{ \sqrt{32\zeta(3) \Sigma_0 }}{\Lambda}\times
\frac{\left(n-1/2\right)}{N} \right]\,.
\end{equation}

At this point the energy spectrum $ E(n)$ given in Eq.~\eqref{eq:En0} still depends on an
inaccurately defined constant $\Sigma_0\approx \log(N/2)$ and its shape only poorly approximates
the exact spectrum plotted in Fig.~\ref{fig:Excitons}. As mentioned above this discrepancy is due
to the fact that the cutoff radius for the sum given in Eq.~\eqref{eq:divsum} is not determined by
the size of the system, i.e $r_c=N/2$, but rather by the shape of the wavefunction $C_n(x)$. To
take into account this mode-dependent cutoff we use the following trick. First, for each
modefunction $C_n(x)$ we replace the constant $\Sigma_0$ in Eq.~\eqref{eq:divsum} by $\Sigma(n)$,
which has a different value for each mode $n$. Second, to derive an accurate expression for
$\Sigma(n)$ we invert the expansion of $C_n(x_j^0)$ outlined in
Eqs.~\eqref{eq:ExcEig3}-\eqref{eq:divsum} and express $\Sigma(n)$ in terms of the full wavefunction
$C_n(x)$ as
\begin{equation}\label{eq:Sigman}
\Sigma(n)=\frac{1}{n(x_i^0)C_n''(x_i^0)}\sum_{j\neq
i}\frac{C_n(x_j^0)-C_n(x_i^0)}{|x_i^0-x_j^0|^3}.
\end{equation}
To evaluate Eq.~\eqref{eq:Sigman} we focus on the region $x_i^0\approx 0$ and neglect small
variations of the density. For the mode functions $C_n(x)$ we insert the harmonic oscillator
functions, $C_n(x)\sim\Phi_{n}(x,\bar \sigma)$, derived above. Around the center of the trap we can
further approximate these wavefunctions by $C_n(x)\sim \cos(k_nx)$ (or $C_n(x)\sim \sin(k_nx)$ for
odd $n$) with a wavevector $k_n^2= (2n-1)/\bar \sigma^2$. For even modes this approximation results
in
\begin{equation}
\Sigma(n)\simeq\frac{1}{k^2_n}\left[2\zeta(3)- \sum_{j\neq 0}\frac{\cos(k_n j)}{|j|^3}\right]\,,
\end{equation}
 which can now be evaluated in a closed form in terms of polylogarithm functions (see
App.~\ref{app:ring}, Eq.~\eqref{eq:J1D}). Expanding the result to lowest order in $k_n$ we finally
obtain
\begin{equation}
\Sigma(n)\simeq \frac{1}{2}\left[3+\log\left( \frac{\bar\sigma^2}{2n-1} \right)\right].
\end{equation}
By replacing $\Sigma_0$ with $\Sigma(n)$ in Eq.~\eqref{eq:En0} we end up with the improved exciton
spectrum
\begin{equation}\label{eq:En}
E(n)=\kappa \left[2\zeta(3)\!-\!A\sqrt{B_N\!+\!\log\left( \frac{N}{n-1/2} \right)}\,
\frac{\left(n-1/2\right)}{N} \right],
\end{equation}
with $B_N=3+\log(\Lambda \sqrt{\log(N/2)/32 \zeta(3)}\,)$ and $A=4\sqrt{\zeta(3)}/\Lambda$. The
corresponding improved modefunctions are given by $C_n(x)\sim\Phi_{n}(x,\sigma_n)$ with a
$n$-dependent width
\begin{equation}
\sigma^2_n= N\frac{\Lambda}{4\sqrt{\zeta(3)}}\left[B_N+ \log\left( \frac{N}{n-1/2}
\right)\right]^{\frac{1}{2}}\,.
\end{equation}
Although our derivation was based on several rather crude approximations we find in comparison with
numerics that our analytic solutions for $E(n)$ and $C_n(x)$ provide an accurate description of the
long wavelength behavior of the exciton spectrum and the shape of the eigenfunctions. Due to the
similarity of the underlying equations this approach should also be applicable to improve the
linear phonon spectrum obtained in Ref.~\cite{Morigi} for a harmonically confined ion crystal.

\emph{Short wavelength limit.} For the short wavelength limit of the exciton spectrum we make the
ansatz $C_n(x_i^0)\equiv C_n(i)=(-1)^i\tilde C_n(x^0_i)$ such that $\tilde C_n(x)$ represents a
slowly varying envelop function for the rapidly oscillating modefunction $C_n(x)$. Inserting this
ansatz into the eigenvalue equation~\eqref{eq:ExcEig} we can proceed in the derivation of $\tilde
C_n(x)$ as explained above for the long wavelength limit. However, in the short wavelength limit
the fast oscillating eigenmodes cancels the effect of long range interactions and in contrast to
the divergent sum in Eq.~\eqref{eq:divsum} the corresponding term in the short wavelength limit has
a well defined value,
\begin{equation}
\frac{1}{2}\sum_{j\neq i}\frac{(-1)^{i-j}}{|x_{i}^0-x_j^0|} \simeq - \log(2) n(x^0_i)\,.
\end{equation}
Therefore, the resulting spectrum in the short wavelength limit is purely harmonic and has the form
\begin{equation}
E(n)=\kappa \left[- \frac{3\zeta(3)}{2} + \frac{\sqrt{24\zeta(3)\log(2)}}{\Lambda}\times
\frac{(\bar n -1/2)}{N}\right],
\end{equation}
with $\bar n = N-n+1$. The envelop functions $\tilde C_n(x)\sim \Phi_{\bar n}(x,\sigma)$ are
harmonic oscillator eigenfunctions with a width
\begin{equation}
\sigma^2=N\Lambda \sqrt{\log(2)/6\zeta(3)}\,.
\end{equation}

\section{Phonon Spectrum of a Harmonically Confined MDC}\label{app:phonons}

In this appendix we derive approximate analytic expressions for the spectrum of longitudinal and
transverse phonons in a harmonically confined MDC. Our calculations are based on a similar approach
used in Ref.~\cite{Morigi} for the phonon spectrum of a 1D ion crystal. However, in contrast to the
ion crystal or the exciton spectrum derived in App.~\ref{app:excitons}, the long range character of
dipole-dipole interactions has no severe effect on the phonon spectrum of a dipolar crystal which
simplifies  calculations. Here we outline only the derivation of the spectrum of longitudinal
phonons. The spectrum of transverse phonons can be derived along the same lines and we give the
results at the end of this appendix.  Note that throughout this appendix we express results in
units of $a(0)=1/n(0)$ and $U_{dd}$ as defined in Sec.~\ref{sec:density}.

To derive the eigenspectrum $\omega(m)$ with $m=1\dots N$ of the Hamiltonian $H_{\rm
phon}^{\parallel}$ given in Eq.~\eqref{eq:HphonPar} we change into the Heisenberg picture where
position operators $x_i(t)$ obey the equation of motion,
\begin{equation}\label{eq:HeisenbergH}
\ddot x_i(t) = - \tilde \nu^2 x_i(t) - \frac{12}{\gamma} \sum_{j\neq i} \frac{ x_i(t)-
x_j(t)}{|x_{i}^0-x_j^0|^5}\,,
\end{equation}
with $\tilde \nu =\hbar \nu/U_{dd}$ the normalized trapping frequency. Using the normal mode
decomposition given in  Eq.~\eqref{eq:NormalModesH},  Eq.~\eqref{eq:HeisenbergH} translates into
the eigenvalue equation,
\begin{equation}\label{eq:EigPhon0}
\gamma\left[\tilde \nu^{2}-\omega^{2}(m)\right]c_{m}(i)=-12\sum_{j\neq
i}\frac{c_{m}(i)-c_{m}(j)}{|x^0_{i}-x^0_{j}|^{5}}\,,
\end{equation}
for the normal modes $c_m(i)$. In the long wavelength limit we can replace the discrete set of
coefficients $c_{m}(x^0_i)\equiv c_{m}(i)$ by a continuous function $c_{m}(x)$ and make the
approximation
\begin{equation}
c_{m}(j)\simeq
c_{m}(i)+c'(x^0_{i})(x^0_{j}-x^0_{i})+c''(x^0_{i})(x^0_{j}-x^0_{i})^{2}/2\,.\end{equation} As the
modefunctions of long wavelength phonons extend over the whole length of the crystal we include
variations of the density $n(x)$ given in Eq.~\eqref{eq:density}, i.e.,
\begin{equation}
x^0_{j}-x^0_{i}\simeq\frac{(j-i)}{n(x^0_{i})}-\frac{(j-i)^{2}}{2}\frac{n'(x^0_{i})}{n^{3}(x^0_{i})}\,.
\end{equation}
Keeping terms up to second order in the derivatives $c_{m}'(x)$ and $n'(x)$ Eq.~\eqref{eq:EigPhon0}
transforms into the differential equation
\begin{equation}
n^{3}(y)c_m''(y)+4n^{2}(y)n'(y)c_m'(y)-\alpha_{m}c_m(y)=0\,,\label{eq:LWdiff}
\end{equation}
with $y=2x/L$ and $\alpha_{m}=2(1-\omega^{2}(m)/\tilde \nu^{2})/3$. Here we made use of the
identity $\gamma\tilde \nu^{2}L^{2}=32\zeta(3)$ (see Sec.~\ref{sec:density}, Eq.~\eqref{eq:Nurel}).
For the density $n(y)=\sqrt[3]{1-y^2}$ we find $n^{2}(y)n'(y)\simeq-2/3y$ and the differential
equation~\eqref{eq:LWdiff} can be solved by the ansatz $c_{m}(y)=\sum_k a_{k}y^{k}$. The
coefficients obey the recursion rule
\begin{eqnarray*} a_{k+2} & = &
a_{k}\left[k(k-1)+8k/3+\alpha_{m}\right]/(k+1)(k+2)\,.\end{eqnarray*}
 The quantization condition for mode $m$, $a_{k+2}=0,\,\forall k>m-1$,
follows from the normalizability of the resulting polynomials and translates into the spectrum
\begin{equation} \omega(m)=\tilde \nu\times\sqrt{1+\left(3m^{2}-m-2\right)/2}\,.\end{equation}
The corresponding modefunctions $c_{m}(x)$ are polynomials extended over the whole length of the
crystal.

In the short wavelength limit we make the ansatz $c_{m}(i)=(-1)^{i}\tilde{c}_{m}(x^0_{i})$ and
repeat the calculations from above for the slowly varying envelop function $\tilde{c}_{m}(x)$. For
$y=2x/L$ we obtain \begin{equation}
\tilde{c}_m''(y)+\frac{4n'(y)}{3n(y)}\,\tilde{c}_m'(y)+\frac{\left(\beta N^2
n^{5}(y)+\alpha_{m}\right)}{n^3(y)}\,\tilde{c}_m(y)=0\,,\label{eq:SWDiff1}
\end{equation}
 with $\beta=31\zeta(5)\Lambda^{2}/24\zeta(3)$ and $\alpha_{m}=8(1-\omega^{2}(m)/\tilde \nu^{2})/9$.
We can simplify Eq.~\eqref{eq:SWDiff1} by neglecting the term proportional to $n'(y)$ and by
expanding the remaining equation up to second order in $y$. This approximation is valid since we
will find below that modefunction of short wavelength phonons are located at the center of the
chain. We end up with
\begin{equation}
\tilde{c}_{m}''(y)+\left[N^2 \beta+\alpha_{m}-(2
N^2\beta/3-\alpha_{m})y^{2}\right]\tilde{c}_{m}(y)=0\,.\end{equation}
Solutions of this equation
are of the form
\begin{equation}\label{eq:SWsigma}
\tilde{c}_{m}(x)\sim H_{\bar m }\left(x/\sigma\right)e^{-x^{2}/2\sigma^2}\,,
\end{equation}
with $\bar m=N-m$ and $\sigma^2=N \lambda^2\sqrt{3/(80\beta)} $. The corresponding spectrum is
given by
\begin{equation}
\omega(m)\simeq\omega_D \left(1-\sqrt{\frac{5}{3\beta} } \frac{(\bar m+1/2)}{N}
\right)\,,\end{equation}  with a Debye frequency $\omega_D=\tilde \nu N \sqrt{9\beta/8}$.

\emph{Transverse phonons.} Along the same lines as shown for the longitudinal phonons we calculate
analytic expressions for the spectrum of transverse phonons determined by the Hamiltonian
$H^\perp_{\rm phon}$ given in Eq.~\eqref{eq:Hphon_perp}. In the long wavelength  limit the
resulting spectrum is
\begin{eqnarray}\label{eq:spec_perpLW}
\omega_{y,z}^\perp(m) \simeq \sqrt{\tilde \nu_\perp^2-\alpha_{y,z}\frac{4\zeta(3)}{\gamma}
\frac{\left(3m^{2}-m-2\right)}{N^2}} \,,
\end{eqnarray}
with $\alpha_y=1$, $\alpha_z=3$ and $\tilde \nu_\perp= \hbar \nu_\perp/U_{dd}$. For the short
wavelength limit we obtain a spectrum of the form
\begin{equation}\label{eq:spec_perpSW}
\omega_{y,z}^\perp(m)\simeq   \sqrt{ \tilde \nu_\perp^2- A\frac{
\,\alpha_{y,z}}{\gamma}\left(1-B\times \frac{(\bar m+1/2)}{N}\right)}\,.
\end{equation}
with numerical constants $A=93\zeta(5)/8\simeq 12.05$ and $B=\sqrt{5\zeta(3)32/31\zeta(5)}\simeq
2.055$.

\section{Lindemann parameter}\label{app:fluc}
In this appendix we calculate the local Lindemann parameter $\Gamma_L(x,T)$ as defined in
Eq.~\eqref{eq:GL} both for the homogeneous and the inhomogeneous crystal. Note that in the
following  we express all quantities in units of $a_0$ and $U_{dd}$. Using the normal mode
decomposition of operators $x_i$ given in Eq.~\eqref{eq:xiAnsatz} for the homogeneous system and in
Eq.~\eqref{eq:NormalModesH} for the inhomogeneous case we obtain
\begin{equation}
\Gamma^2_{L}(x_i^0,T)=\sum_{k}\frac{n(x_i^0)}{2\gamma\omega(k)}|c_k(i+1)-c_k(i)|^{2}(2N(\omega(k))+1)\,,\end{equation}
with  $N(\omega(k))$ the thermal occupation number  of mode $k$. For a homogenous system $k$ is the
quasi momentum, $c_k(i)$ are plane waves and $\omega(k)=f(k)/\sqrt{\gamma}$. In the homogeneous
system we replace $k$ by the index $m=1,\dots N$ and the normal modes $c_m(i)$ with the
corresponding spectrum $\omega(m)$ are discussed in Sec.~\ref{sec:Harmonic} and
App.~\ref{app:phonons}.  For simplicity we adopt the sound wave approximation $\omega(m)\simeq
\alpha_D/\sqrt{\gamma}\times m/N$ with $\alpha_D\approx 6.95$. The Lindemann parameter then has the
general form
\begin{equation}
\Gamma_{L}(x,T)=\frac{1}{\gamma^{\frac{1}{4}}} \times F \left(\xi=2x/L,\tau =\sqrt{\gamma}\,
T\right)\,.\end{equation} For the homogeneous crystal $F(\xi,\tau)=F_h(\tau)$ with
\begin{equation}
F_h^2\left(\tau\right)=\frac{2}{\pi}\int_{0}^{\pi}\frac{dk}{f(k)}\sin^{2}(k/2)\left(\frac{2}{e^{f(k)/\tau}-1}+1\right)\,.
\end{equation}
In the two limits of interest the numerical values of this function are $F_h(\tau \rightarrow
0)\simeq 0.424$, and $F_h(\tau \gg 1)\simeq 0.278 \times \sqrt{\tau}$. In the inhomogeneous case we
obtain
\begin{equation}\label{eq:F}
F^2\left(\xi, \tau\right)=\frac{2}{N\alpha_D\Lambda^2 }  \sum_m
\frac{\left[c_m'(\xi)\right]^2}{n(\xi) \,m }\!\left(\frac{2}{e^{\alpha_D m /(\tau
N)}-1}+1\right)\,.
\end{equation}
Numerical values of this function are plotted in Fig.~\ref{fig:Fluctuations}.




\bibliographystyle{apsrev}

\end{document}